\documentclass[aps,pra,twocolumn,longbibliography,amsfonts, amssymb, amsmath, amsthm,groupedaddress]{revtex4-2}
\usepackage{url}
\usepackage{color}
\usepackage{soul} 
\usepackage{dcolumn}
\usepackage{bm}
\usepackage{physics}
\usepackage{amsmath}
\usepackage[matrix,frame,arrow]{xypic}
\usepackage{comment}
\usepackage{qcircuit}
\usepackage{graphicx}
\usepackage{verbatim}           
\usepackage[utf8]{inputenc}
\usepackage{appendix}
\usepackage[colorlinks=true,linkcolor=blue,urlcolor=blue,citecolor=blue]{hyperref}
\usepackage{caption}
\usepackage{subcaption}

\usepackage{amsthm}             
\theoremstyle{plain}            

\def\bra#1{{\langle#1|}}
\def\ket#1{{|#1\rangle}}

\def\phit{{\ket{\phi_\theta}}}
\def\phitbar{{\ket{\bar{\phi}_\theta}}}
\def\braphit{{\bra{\phi_\theta}}}
\def\braphitbar{{\bra{\bar{\phi}_\theta}}}

\begin{document}

\title{Weakly Fault-Tolerant Computation in a Quantum Error-Detecting Code}
\author{Christopher Gerhard}
\email{cgerhard@usc.edu}
\author{Todd A. Brun}
\email{tbrun@usc.edu}
\affiliation{Ming Hsieh Department of Electrical and Computer Engineering, \\
University of Southern California, Los Angeles, California}

\begin{abstract}
Many current quantum error-correcting codes that achieve full fault tolerance suffer from having low ratios of logical to physical qubits and significant overhead. This makes them difficult to implement on current noisy intermediate-scale quantum (NISQ) computers and results in the inability to perform quantum algorithms at useful scales with near-term quantum processors. As a result, calculations are generally done without encoding. We propose a middle ground between these two approaches: constructions in the $[[n,n-2,2]]$ quantum error-detecting code that can detect any error from a single faulty gate by measuring the stabilizer generators of the code and additional ancillas at the end of the computation. This achieves {\it weak fault tolerance}. As we show, this yields a significant improvement over no error correction for small computations with low enough physical error probabilities and requires much less overhead than codes that achieve full fault tolerance. We give constructions for a set of gates that achieve universal quantum computation in this error-detecting code, while satisfying weak fault tolerance up to analog imprecision on the physical rotation gate.
\end{abstract}

\date{May 21, 2026}

\maketitle

\section{Introduction}

One of the main obstacles to implementing quantum algorithms in real-world systems is their susceptibility to noise. To combat this, a rich theory of quantum error correction (QEC) has been developed to achieve fault-tolerant quantum computation (FTQC) \cite{DiVincenzo_1996,gottesman2009introduction,Yoder_2016,nielsen_chuang_2010,lidar_brun_2013,Knill_2000,Fowler_2012}. However, these codes typically introduce significant overhead into any computation. This arises from a number of causes, especially the low rates of most codes suitable for FTQC and the Eastin-Knill no-go theorem proving that no code allows universal quantum computation through only transversal gate operations \cite{Eastin_2009}. As a result, many fault-tolerant schemes employ magic state distillation to achieve fault-tolerant non-Clifford gates \cite{Bravyi_2005}. New methods have been introduced in the past few years to reduce the cost of this procedure \cite{Litinski_2019,Fowler_2013,Gidney_2019,Bravyi_2012} and improve the error rates of the initial magic states themselves \cite{gidney2023cleanermagicstateshook}, but magic state distillation still considerably magnifies the size of the logical versions of quantum circuits. The resulting overhead puts these codes out of reach of current or near-term noisy intermediate-scale quantum (NISQ) computers, since state-of-the-art general quantum computers have at most a few hundred qubits. Use of these codes can limit the user to a very small number of logical qubits, since one often needs to encode logical qubits in hundreds or thousands of physical qubits to achieve full fault tolerance \cite{brun2015teleportationbased}. As a result, FTQC at interesting scales is difficult for these systems and must await the future development of much larger, more capable quantum computers.

Significant efforts have been made to reduce the resource requirements and overhead for fully fault-tolerant quantum error correction. One promising direction is known as flag fault tolerance, which has been demonstrated in a number of papers for common distance $3$ codes \cite{Chao_2018,Chao_2018b,Yoder_2017}. In flag fault tolerance, ancilla qubits are used as flags to signal the presence of uncorrectable logical errors in a quantum circuit. As an example, the authors of \cite{Chao_2018} reduced the qubit requirements of fault-tolerant quantum error correction by using only 2 ancillas for the $[[5,1,3]]$, $[[7,1,3]]$, and $[[15,7,3]]$ codes. Extensions of this scheme to arbitrary distance codes were found by imposing a set of conditions on the code family \cite{Chamberland_2018}. Following this paper, Chao and Reichardt were able to extend the flag fault tolerance scheme unconditionally to any stabilizer code \cite{Chao_2020}. Their flag-based methods inspired the approach we take in this paper to achieve error detection. Further resource reductions in flag fault-tolerant circuits have been demonstrated in a number of recent papers \cite{Bhatnagar_2023,Tansuwannont_2023,anker2024flaggadgetsbasedclassical,liou2024reducingquantumerrorcorrection}.

Flag fault tolerance schemes have a very small qubit overhead, but they do require dynamical circuits where each potential circuit can be relatively long \cite{heußen2024efficientfaulttolerantcodeswitching}. One way to get around this issue at the cost of a potentially larger qubit overhead is to use a complete set of transversal logic gates through code switching \cite{Anderson_2014,gupta2024universaltransversalgates,Butt_2024}. Promising work to efficiently implement such a universal fault-tolerant scheme was carried out in a number of recent papers \cite{heußen2024efficientfaulttolerantcodeswitching,Butt_2024,butt2024measurementfree}. There have also been efforts to go beyond the traditional code-centered view of quantum error correction and instead take a circuit-centered view, which focuses on codes generated from the quantum circuit itself. Explicitly, this formalism relies on the observation that the set of all output bit strings of a Clifford circuit is actually itself a linear code. For a more complete treatment, see \cite{delfosse2023spacetimecodescliffordcircuits}. Many elements of fault tolerance have also been demonstrated experimentally in the past decade \cite{Linke_2017,Andersen_2020,Postler_2024,Egan_2021,Anderson_2021,Abobeih_2022,Krinner_2022,Zhao_2022,Google_2023,Bluvstein_2023,ryananderson2024,mayer2024,pogorelov2024,reichardt2024demonstrationquantumcomputationerror,Hong_2024,paetznick2024}.

A potential compromise for small computations, which maintains some level of error protection while increasing the qubit overhead only modestly, is to only detect errors instead of correcting them. A code of distance $d$ can correct any error up to weight $(d-1)/2$, but can {\it detect} any error up to weight $d-1$. However, increasing the qubit rate in this way incurs the penalty that we no longer know exactly what type of error occurred. One can avoid having to correct errors in relatively small computations using postselection, where one only keeps the computational runs where no errors were detected. Since most computations on NISQ computers are of relatively small size, the extra overhead incurred from this is usually not too bad, and the effects of errors are reduced by eliminating erroneous runs. However, even with the less stringent requirement of only detecting errors, the overhead incurred from most current methods is still significant. 

This motivates the need for a scheme that does not incur too much overhead for current quantum computers, so that interesting calculations can still be carried out with at least some protection. One such approach in a recent paper \cite{delfosse2024lowcostnoisereductionclifford} uses gate teleportation for Clifford circuits. The authors' Clifford noise reduction (CliNR) scheme results in a reduction in the overhead associated with gate teleportation by performing a smaller set of random stabilizer measurements to detect errors in the resource states consumed. Offline fault-detection of errors in the resource states is implemented to improve scalability. Importantly, CliNR achieves a vanishing logical error rate in the regime where the physical error rate $p$ goes to zero; strictly, the logical error rate goes to zero for circuits of size $s=O(1/p^2)$ when $sp\rightarrow{0}$ and $nsp^{2}\rightarrow{0}$. Here, $n$ is the number of physical qubits; this will be our convention for the rest of the paper. We will also use $p$ to represent the physical error rate unless otherwise stated. In contrast, the direct implementation without CliNR only achieves a vanishing logical error rate if $s=O(1/p)$. Another promising proposal that achieves partial fault tolerance involves a general architecture for early fault-tolerant quantum computing know as the ``space-time efficient analog rotation quantum computing'' (STAR) architecture \cite{Akahoshi_2024,akahoshi2024compilationtrotterbasedtimeevolution,toshio2024practicalquantumadvantagepartially}. The scheme outlined in this set of papers involves fault-tolerant Clifford operations implemented via lattice surgery in surface codes, while non-Clifford gates are implemented by gate teleportation. In preparing the required resource states, error mitigation and control error cancellation techniques are used to greatly reduce the probability of errors. In addition, optimal postselection strategies are implemented to increase the probability of success in preparing the required resource state. In a direction more similar to our own approach, recent work involving low overhead quantum error detection has been done for trapped-ion computers in \cite{self2022protecting} and \cite{Yamamoto_2024}. These papers are especially interesting to us because the authors of \cite{self2022protecting} tailored the $[[6,4,2]]$ quantum error-detecting code (QEDC) to the underlying quantum hardware and then implemented it experimentally on a Quantinuum trapped-ion computer, while the authors of \cite{Yamamoto_2024} applied the results of \cite{self2022protecting} to a particular problem in quantum chemistry. This approach is similar to the one we have taken in this paper. The main difference is that we tailor the logical operations on the $[[n,n-2,2]]$ code to achieve what we call weak fault tolerance, which we will rigorously define below. Initial work on finding fault-tolerant operations in the $[[n,n-2,2]]$ code was carried out by Gottesman \cite{Gottesman_1998}. Building on this, Chao and Reichardt developed a set of flag-based fault-tolerant encoded gates \cite{Chao_2018b}. Their fault-tolerant encoded Clifford gates include a targeted Hadamard and a CZ gate. In contrast to their scheme, where they measure flags at the end of a gadget, we introduce a protocol that defers all measurements to the end of a quantum computation to avoid conditional operations, and introduce a new set of weakly fault-tolerant Clifford gates. Our gates are $R_{\rm ZZ}$ and $R_{\rm XX}$ rotations with a fixed absolute angle of $\frac{\pi}{2}$. These M{\o}lmer-S{\o}rensen-type rotations are natural operations for ion trap quantum computers and are straightforward to implement in these architectures \cite{S_rensen_1999, self2022protecting}; in other architectures, they can be implemented in terms of other two-qubit gates. Although these rotation operators are well-known, we give novel flag-based constructions that make them weakly fault-tolerant.

This paper introduces a set of encoded gates in the $[[n,n-2,2]]$ QEDC that, together, allow universal quantum computation, while detecting any single gate error during the computation, up to analog errors on our rotation gates. An analog error refers to imprecision in our rotation gate that results in a valid rotation by an amount different from the one we wanted. Aside from such errors, our scheme allows the detection of any error on our Clifford gates up to first order in the probability of a single gate error; many higher-order errors are also detectable, but not all. Here we treat errors on different gates as uncorrelated and equally likely, for simplicity. Errors are detected by measuring the stabilizer generators of the QEDC and of the additional ancillas used in the gate constructions. This scheme is especially beneficial for current and near-term NISQ machines because it reduces the probability of undetectable errors in a modest-sized computation by an order of magnitude for a low-enough physical error rate when compared with the case of no encoding. Importantly, our weakly fault-tolerant logical gate set does not spoil the high rate of the QEDC. The ancillas used for added protection can be reused for all gates, so this construction is resource efficient. Since computations on NISQ machines are of relatively small size, the loss in rate due to discarded runs from errors should not be too bad.

\section{The $[[n,n-2,2]]$ QEDC and weak fault tolerance}
\label{sec:QEDC and weak fault tolerance}

One of the simplest ways to add some level of error protection to a quantum computer is to designate two qubits as parity checks for all of the other qubits. This can be seen as a direct analogue of the classical parity-check code, which only has a single parity check bit to detect bit-flip errors. In its quantum extension, we require a second parity-check qubit to detect phase flip errors. Conventionally, if we have $n$ total qubits, then the $(n-1)$th and $n$th qubits keep track of the Pauli $X$ and $Z$ operator parities, respectively. We require that $n$ be even, so that the $X$ and $Z$ parities commute with each other, and an odd parity indicates an error. The stabilizer generators of this code are the all-$X$ and all-$Z$ operators on the physical qubits. For example, in the $[[4,2,2]]$ code the stabilizer generators are $XXXX$ and $ZZZZ$. This code can only detect errors, since the minimum distance required for a code to correct a single error is 3. Despite this, the code is still useful for small computations, because we can repeat a computation until we do not detect any errors. If a computation is run multiple times (as is generally the case), then this procedure is just postselection on not detecting an error.

To use the $[[n,n-2,2]]$ QEDC for quantum algorithms, one must implement logical operations on the encoded qubits. We will mostly use a standard encoding, in which we associate the $i$th logical qubit with the $i$th physical qubit in the code for $i=1,\ldots,n-2$. Any logical operation must commute with the stabilizer generators of the code and implement a non-trivial operation on the encoded quantum state. We need $n-2$ distinct logical $X$ and $Z$ operators, since we have $n-2$ logical qubits. The logical $X$ operator on the $i$th logical qubit is a weight-two Pauli $X$ operator on the $i$th physical qubit and the $X$ parity check qubit, that is, $\bar{X}_i \equiv X_i X_{n-1}$. The logical $Z$ operator is defined similarly using the $Z$ parity check qubit and the $i$th $Z$ operator, $\bar{Z}_i \equiv Z_i Z_{n}$. A logical $Y$ operator is just the product of the prior two operators (up to a phase), $\bar{Y} = i \bar{X}\bar{Z}$. This completely specifies a basis for the QEDC.

We now need a rigorous definition for weak fault tolerance. In this paper, weak fault tolerance means that any error produced by a single faulty gate is transformed to a detectable error by the end of the computation. An error is considered detectable if it anticommutes with one of the stabilizer generators of the QEDC or any of the additional ancilla stabilizers. This will cause the parity to flip and indicate that an error occurred. Any valid logical operation will commute with all the stabilizers. We model errors by depolarizing noise: we assume that a faulty gate is equivalent to the correct gate followed by any possible Pauli error on the qubits it operated on. These operators will be described more precisely when we introduce the symplectic formalism in the next section. Weak fault tolerance can also be considered error detection up to first order in our error model if we assume that errors on different gates are independent. As a result, any scheme that achieves weak fault tolerance essentially reduces the probability of an undetectable error to $O(p^2)$. And of course, many errors comprising multiple faults will also be detectable.

From a mathematical point of view, we define a weakly fault tolerant protocol as one in which the probability of an undetectable error at the end of a quantum circuit scales as $O(p^2)$ in the physical error rate $p$. Such a protocol can achieve this scaling through measurements and postselection at the end of a computation. Weakly fault-tolerant protocols are not scalable on their own, and suffer from $O(g^2p^2)$ scaling, where $g$ is the number of noisy gates in a quantum circuit. This is more severe in weakly fault-tolerant protocols, since pairs of faults can occur in different quantum circuit gadgets to produce undetectable errors (even though these are $O(p^2)$ events). A circuit gadget is a building block that implements some specific function, such as a gate gadget or a measurement gadget. This scaling is different from fully fault-tolerant protocols, where errors at the end of a fault-tolerant gadget are acted upon to prevent them from spreading to other blocks of the code. Due to this containment of errors, undetectable faults can only occur due to multiple errors within a single gadget or on pairs of gates between fault-tolerant gadgets. Many of these types of $O(g^2)$ faults will still be detectable in a weakly fault-tolerant protocol, so their main impact will be on the postselection rate of the complete quantum computation. This is precisely our main justification for terming such a protocol {\it weakly} fault-tolerant. It can achieve elements of fault tolerance with a lower overhead than fully fault-tolerant protocols, but at the cost of scalability.

In this paper, we consider weak fault tolerance in the case where errors on different physical gates are independent and do not include correlated errors, such as crosstalk, that occur in real quantum devices. If we narrow our view to the gadget level, our definition of weak fault tolerance is equivalent to the standard definition of a fault-tolerant gate gadget by Gottesman \cite{gottesman2009introduction} if we include postselection. Based on this local definition of fault tolerance, the weakly fault-tolerant Clifford circuits we present in this paper can be considered fault-tolerant. However, our gate constructions are flag-based. In traditional flag fault-tolerant schemes like those in \cite{Yoder_2017,Chao_2018b,Chamberland_2018,delfosse2020,Tansuwannont2020,Reichardt_2020}, the flag ancillas are measured at the end of a gadget to catch errors. These measurement results can then be used to implement conditional operations to act on any raised flags. Since we only measure the flag ancillas at the end of a computation, we are omitting a gadget-level step that is crucial to fully fault-tolerant flag-based gadgets. As such, we refer to all of our Clifford gate gadgets as weakly fault-tolerant, to explicitly acknowledge the lack of this mid-circuit measurement.

In contrast to early fault tolerance (EFT) as defined in \cite{Katabarwa_2024,Liang_2024,Kshirsagar_2024}, weak fault tolerance is not equivalent to full fault tolerance at any scale beyond the Gottesman gate-gadget level. The reason for this departure is an attempt to reduce the overhead incurred from our protocol even further, so that it can be implemented on current and near-term quantum devices. As we will show, this comes at the potential cost of having to repeat the circuit a large number of times. The same idea of trading off overhead for an increase in the number of circuit runs required to perform a computation is present in some current EFT algorithms \cite{Katabarwa_2024}. However, our notion of weak fault tolerance is more similar in spirit to the CliNR algorithm outlined in \cite{delfosse2024lowcostnoisereductionclifford}. Their protocol and the one we outline in this paper are not fault-tolerant. However, both protocols reduce the logical error rate for certain values of $p$ while incurring a relatively small amount of overhead compared to fully fault-tolerant schemes. In contrast to \cite{delfosse2024lowcostnoisereductionclifford}, weak fault tolerance requires that the probability of a logical error is of $O(p^2)$ for any value of $p$ after postselection on the final stabilizer measurements.

For non-Clifford gates we will consider encoded rotations. In this case, we generally cannot detect {\it analog} errors that are equivalent to an error in the rotation angle. We will discuss how one can go beyond this limitation, at least in principle, but it generally demands greater resources that may be impractical in small, near-term quantum computations. We will also argue that as the capabilities of quantum computers increase, weak fault tolerance can be strengthened to eventually encompass fully fault-tolerant operation, which is required for truly scalable quantum computing.

\section{Encoded Clifford Gates}

To use the $[[n,n-2,2]]$ QEDC for quantum computation, we need to find a set of logical gates that can implement any quantum operation. It is useful to decompose this set into Clifford and non-Clifford operations, since the latter are typically much more difficult to do in a stabilizer code. This section will focus solely on Clifford operations, and we will consider non-Clifford operations in the next section.

Any valid encoded Clifford gate must leave the codespace of the QEDC invariant, so we will only consider operations that commute with its stabilizer generators. As one can easily check, there are no non-trivial one-qubit operations that do this; $X_i$, $Y_i$, and $Z_i$ all flip the phase of one or both generators, where the subscript $i$ indicates an operation on the $i$th qubit, and any single-qubit unitary is a linear combination of these operators and the identity. These are detectable errors and are not valid operations in this code. If we instead look at 2-qubit gates, there exist Clifford operations that commute with all of the stabilizer generators and implement non-trivial logical operations on the encoded qubits. Some examples are the $\exp(i\theta X_{i}X_{j}/2)$, $\exp(-i\theta Y_{i}Y_{j}/2)$, and $\exp(-i\theta Z_{i}Z_{j}/2)$ gates with an absolute fixed angle of $\frac{\pi}{2}$ described below. All three commute with the all-$X$ and all-$Z$ stabilizers on the physical qubits, yet they act nontrivially on the logical operators. These gates themselves are not weakly fault-tolerant because they can lead to weight $2$ errors that commute with all the stabilizer generators. However, we will show that these undetectable errors can be removed to first order by introducing two additional ancillas and using a slightly more complex circuit. We can then use these 2-qubit operations to construct the weakly fault-tolerant encoded CNOT, Phase, and Hadamard gates, which generate all encoded Clifford operations \cite{Bravyi_2005}.

\subsection{The gate set}

In the $[[n,n-2,2]]$ QEDC, one can generate any 2-qubit Clifford operation that commutes with the stabilizer generators from a set of three quantum gates. We will prove this using the binary symplectic representation. This formalism makes it straightforward to show that these gates arise as solutions to a set of linear equations. We call these the SWAP, $R_{\rm ZZ}$, and $R_{\rm XX}$ gates. The SWAP gate is the most straightforward of the three: it swaps two of the physical qubits in our code. In architectures where quantum operations are not limited by physical distance, one can implement this gate by simply relabeling the qubits, making it error-free. Examples of architectures that feature all-to-all connectivity and/or high-fidelity swap gates include shuttling-based ion traps and recent implementations of neutral atom quantum computers. For a complete discussion of these architectures, see \cite{schoenberger2024shuttlingscalabletrappedionquantum}, \cite{Sun_2024}, and \cite{reichardt2024logicalcomputationdemonstratedneutral}.

The other two gates in our set are 2-qubit rotation gates. The $R_{\rm ZZ}$ gate is the unitary operator
\begin{equation} 
R_{\rm ZZ} = \frac{1}{\sqrt{2}}(I - iZZ) ,
\label{ZZ Unitary}
\end{equation}
where $i$ is the imaginary unit and $ZZ$ is the product of Pauli Z operators on the two qubits. Similarly, the $R_{\rm XX}$ gate is the unitary
\begin{equation}
R_{\rm XX} = \frac{1}{\sqrt{2}}(I + iXX) .
\label{XX Unitary}
\end{equation}
As we will show shortly, these unitary operators are equivalent to our binary symplectic versions of the $R_{\rm XX}$ and $R_{\rm ZZ}$ gates, up to a phase. We will consider and correct for this phase later in this section.

Representing these gates as unitary matrices is cumbersome for large circuits, where a unitary on $n$ qubits is a $2^n \times2^n$ matrix. We will instead use their binary symplectic representations in our analysis, which we now introduce.

\subsection{Binary symplectic representation}

Although some readers may already be familiar with the binary symplectic representation, we briefly introduce it for completeness. For a more complete treatment, see \cite{Calderbank_1997}, \cite{PhysRevA.68.042318} and \cite{Rengaswamy_2020}. An element $g$ of the $n$-qubit Pauli group $\mathcal{G}_n$ can be written in the form
\begin{equation}
g \in \mathcal{G}_n = (i)^\ell Z^{\underline{z}} X^{\underline{x}} = (i)^\ell Z^{z_1} X^{x_1} \otimes ... \otimes Z^{z_n} X^{x_n} ,
\end{equation}
where $i$ is the imaginary unit, $\ell \in \{0,1,2,3\}$, $\underline{x}$ and $\underline{z}$ are binary $n$-vectors $(x_1x_2\cdots x_n)$ and $(z_1z_2\cdots z_n)$, respectively and their elements $\{x_j\}$ and $\{z_j\}$ are $0$ or $1$ (that is, they are bits). We represent $g$, up to a phase, by the {\it symplectic vector} $\underline{g}$:
\begin{equation}
g \rightarrow \underline{g} = (\underline{x} | \underline{z}) = (x_1 x_2 \cdots x_n | z_1 z_2 \cdots z_n) .
\end{equation}
We do not specify the phase $(i)^\ell$ in this representation, but we can keep track of it separately if we wish, and the phase has no effect on the commutation properties of the operators. For now we will ignore that phase, which can be adjusted afterwards (as we will show).

Two operators $g$ and $g'$ with vectors $\underline{g}$ and $\underline{g}'$ commute if their {\it symplectic inner product} $\underline{g} \odot \underline{g}'$ is zero. We define the symplectic inner product as
\begin{equation}
\underline{g} \odot \underline{g}' = \underline{x} \vdot \underline{z'} + \underline{z} \vdot \underline{x'} = \sum_{i=1}^{n} x_i z_i' + z_i x_i' ,
\end{equation}
where all arithmetic is binary (so the sum is modulo 2). The two operators anticommute if $\underline{g} \odot \underline{g}' = 1$.

We can represent a complete set of $2n$ generators, which is the set of all single-qubit operators, as a $2n\times 2n$ matrix,
\[
\underline{M} = \begin{pmatrix} \underline{M_x} & | & \underline{M_z} \end{pmatrix} ,
\]
where each row represents one generator. $\underline{M}_x$ and $\underline{M}_z$ are themselves matrices with $n$ columns and $2n$ rows. The canonical set of generators are just the standard binary basis vectors. With this choice we have $\underline{M} = \underline{I}$
where $\underline{I}$ is the identity matrix. The rows of this matrix represent $2n$ operators $X_1,X_2,...,X_n,Z_1,Z_2,...,Z_n$. They represent $n$ pairs of anticommuting operators: $(X_1,Z_1),(X_2,Z_2),...,(X_n,Z_n)$. The operators in each pair anticommute with each other, but commute with all of the other pairs. These pairs are called {\it symplectic partners}. In this paper we will only work with sets of generators that form $n$ pairs of symplectic partners.

We can also define a symplectic matrix $\underline{J}$:
\begin{equation}
\underline{J} = \begin{pmatrix} \underline{0}_{n\times n} & \underline{I}_{n\times n} \\ \underline{I}_{n\times n} & \underline{0}_{n\times n} \end{pmatrix} ,
\end{equation}
where the subscript $n\times n$ denotes an $n$-by-$n$ block in the $\underline{J}$ matrix. With the matrix $\underline{J}$ we can write the symplectic inner product as 
\[
\underline{g} \odot \underline{g}' = \underline{g} \underline{J} (\underline{g}')^T .
\]
Additionally, we can always order our set of generators, which is represented by the rows of $\underline{M}$, so that
\begin{equation}
\underline{M} \underline{J} \underline{M}^T = \underline{J}
\label{eq:symplecticAnticommutation}
\end{equation}
holds. These linear equations represent the canonical anticommutation relations, which should always be preserved by unitary transformations.

We are now ready to describe Clifford operations. In their binary symplectic form, Clifford operators can be represented as $2n\times 2n$ binary matrices. They transform binary symplectic row vectors through column operations:
\begin{equation}
\underline{g} \rightarrow \underline{g}' = \underline{g} \underline{C} ,
\end{equation}
where $\underline{C}$ represents a Clifford operator. In a similar way, a Clifford operator can transform an entire set of generators by
\begin{equation}
\underline{M} \rightarrow \underline{M}' = \underline{M}\underline{C} .
\label{eq:cliffordSymplecticTransformation}
\end{equation}
Since they are unitary, Clifford operators preserve the anticommutation relations in Eq~(\ref{eq:symplecticAnticommutation}):
\begin{equation}
\underline{C} \underline{J} \underline{C}^T = \underline{J} .
\label{eq:CliffordAnticommutation}
\end{equation}

This completes our introduction of the binary symplectic formalism and will allow us to analyze encoded Clifford operations.

\subsection{Encoded Clifford Gates}

We are now ready to prove that the entire encoded Clifford group in the $[[n,n-2,2]]$ QEDC can be generated by the SWAP, $R_{\rm XX}$, and $R_{\rm ZZ}$ gates. Our approach to this will be very similar to the general algorithm outlined by Rengaswamy et al. \cite{Rengaswamy_2020} for synthesizing the logical Clifford operators of stabilizer codes. We begin by first imposing a set of restrictions that translate to our encoded operators belonging to the Clifford group and leaving the codespace invariant. These conditions can be written compactly as a set of binary symplectic matrix equations. One solution set to these equations are the encoded SWAP, $R_{\rm XX}$, and $R_{\rm ZZ}$ gates. It is important to note that this set is not a unique solution, and there are likely other gate decompositions that generate all encoded Clifford operations.

We start by deriving a binary symplectic matrix equation that completely determines the form that our encoded Clifford operations can take. One restriction on what the encoded Clifford group can look like in the QEDC is imposed by Eq.~\eqref{eq:CliffordAnticommutation}. In addition, we require that these operations leave the stabilizer generators of the code unchanged. We focus on 2-qubit Clifford operations. In the binary symplectic formalism, this restriction translates to the linear equations
\begin{subequations}
\begin{eqnarray}
\begin{pmatrix} 1 & 1 & 0 & 0 \end{pmatrix} \underline{C} &=& \begin{pmatrix} 1 & 1 & 0 & 0 \end{pmatrix} , \\
\begin{pmatrix} 0 & 0 & 1 & 1 \end{pmatrix} \underline{C} &=& \begin{pmatrix} 0 & 0 & 1 & 1 \end{pmatrix} ,
\end{eqnarray}
\end{subequations}
where the Clifford operator $\underline{C}$ is a $4\times4$ matrix. By writing $\underline{C}$ in block form and imposing the anticommutation relations in Eq.~(\ref{eq:CliffordAnticommutation}), we get a set of linear equations. There are eight solutions to these equations that are generated by three matrices: the SWAP, $R_{\rm XX}$, and $R_{\rm ZZ}$ gates. All other solutions are products of these three. Their binary symplectic representations are:
\begin{equation}
\begin{aligned}
\underline{C}_{\rm SWAP} &= \begin{pmatrix} 0 & 1 & 0 & 0 \\ 1 & 0 & 0 & 0 \\ 0 & 0 & 0 & 1 \\ 0 & 0 & 1 & 0 \end{pmatrix} , \qquad
\underline{C}_{\rm R_{\rm ZZ}} = \begin{pmatrix} 1 & 0 & 1 & 1 \\ 0 & 1 & 1 & 1 \\ 0 & 0 & 1 & 0 \\ 0 & 0 & 0 & 1 \end{pmatrix} , \\[1em]
\underline{C}_{\rm R_{\rm XX}} &= \begin{pmatrix} 1 & 0 & 0 & 0 \\ 0 & 1 & 0 & 0 \\ 1 & 1 & 1 & 0 \\ 1 & 1 & 0 & 1 \end{pmatrix} .
\label{eq:binarySymplecticQEDCGates}
\end{aligned}
\end{equation}
This gives us a set of three gates that can implement any 2-qubit Clifford transformation that leaves the stabilizer generators of the QEDC unchanged.

We can now prove that these three gates are sufficient to implement any encoded Clifford operation on the $[[n,n-2,2]]$ QEDC. We will refer to this set of operations as the encoded Clifford group. A straightforward way to show that these gates generate all encoded Clifford operations is to build the encoded CNOT, Phase, and Hadamard gates from this set. Since the CNOT, Phase, and Hadamard gates can be used to construct any Clifford circuit, the same will also be true for their logical versions in the QEDC. Figs.~\ref{fig:CNOT}--\ref{fig:Hadamard} show how to construct these gates, up to single-Pauli corrections, on the $[[4,2,2]]$ and $[[n,n-2,2]]$ QEDC. Specifically, the circuits depicted involve the unitary versions of the $R_{\rm XX}$ and $R_{\rm ZZ}$ gate given in Eqs.~\eqref{ZZ Unitary} and \eqref{XX Unitary}. These circuits are derived from the binary symplectic versions of these two gates, which do not consider the phases; therefore, these constructions can introduce phase errors, which can be corrected via single-qubit Pauli operations, which we will demonstrate in the next section. As a result, the SWAP, $R_{\rm ZZ}$, and $R_{\rm XX}$ gates are sufficient to generate the encoded Clifford group in the QEDC. Further details on these circuits are provided in Appendix~\ref{appendix:nonWeaklyFaultTolerantCircuitWalkthrough}.

It is important to note that the $[[4,2,2]]$ code differs from the general $[[n,n-2,2]]$ QEDC. It has a valid set of logical operators 
\begin{eqnarray}
{X_1 , Z_1} &=& {XXII , IZZI} \nonumber\\
{X_2 , Z_2} &=& {IXXI , ZZII} ,
\label{eq:422logops}
\end{eqnarray}
where X and Z are once again our Pauli operators. One can show that this is a valid set of logical operators for the $[[4,2,2]]$ QEDC. Because of this simple form we can implement a logical CNOT by a single SWAP gate. For higher numbers of qubits $n>4$ a more complicated circuit is required, as Fig.~\ref{fig:CNOT} shows.

\begin{figure*}[tbp]
    \centering
    \begin{subfigure}{0.36\textwidth}
        \centering
        $\Qcircuit @C=7em @R=1em {
            \lstick{1} & \qw & \qw & \\
            \lstick{2} & \qswap \qwx[1] & \qw & \\
            \lstick{3} & \qswap & \qw & \\
            \lstick{4} & \qw & \qw & 
        }$
        \caption{}
    \end{subfigure}
    \begin{subfigure}{0.45\textwidth}
        \centering
        $\Qcircuit @C=7em @R=1em {
            \lstick{1} & \qswap \qwx[1] & \qw \\
            \lstick{2} & \qswap & \qw \\
            \lstick{3} & \qw & \qw \\
            \lstick{4} & \qw & \qw 
        }$
        \caption{}
    \end{subfigure}
    $\vspace{0.5cm}$
    \begin{subfigure}{0.9\textwidth}
        \centering
        $\Qcircuit @C=3em @R=1.8em {
            \lstick{j} & \qw & \qw & \control \qw \ar @{-} [d] ^{\raisebox{0.6ex}{$R_{\rm ZZ}$}} \qwx[3] & \qw & \qw & \control \qw \ar @{-} [d] ^{\raisebox{0.6ex}{$R_{\rm ZZ}$}} \qwx[3] & \qw & \qw \\
            \lstick{k} & \control \qw \ar @{-} [d] ^{\raisebox{0.6ex}{$R_{\rm XX}$}} \qwx[2] & \qw & \qw & \qw & \control \qw \ar @{-} [d] ^{\raisebox{0.6ex}{$R_{\rm XX}$}} \qwx[2] & \qw & \control \qw \ar @{-} [d] ^{\raisebox{0.6ex}{$R_{\rm XX}$}} \qwx[1] & \qw \\
            \lstick{n-1} & \qw & \control \qw \ar @{-} [d] ^{\raisebox{0.6ex}{$R_{\rm XX}$}} \qwx[1] & \qw & \control \qw \ar @{-} [d] ^{\raisebox{0.6ex}{$R_{\rm XX}$}} \qwx[1] & \qw & \qw & \control \qw & \qw \\
            \lstick{n} & \control \qw & \control \qw & \control \qw & \control \qw & \control \qw & \control \qw & \qw & \qw
        }$
        \caption{}
    \end{subfigure}
    \caption{Circuits for the logical CNOT gate in the $[[4,2,2]]$ and $[[n,n-2,2]]$ QEDCs. $R_{\rm ZZ}$ and $R_{\rm XX}$ rotation gates have a fixed angle of $+\frac{\pi}{2}$ and $-\frac{\pi}{2}$, respectively. Circuits (a) and (b) show the CNOT from logical qubit 1 to logical qubit 2 and logical qubit 2 to logical qubit 1, respectively. Circuit (c) performs a CNOT from logical qubit $j$ to logical qubit $k$ in the $[[n,n-2,2]]$ code for $n>4$.}
    \label{fig:CNOT}
\end{figure*}

\begin{figure*}[tbp]
    \centering
    \begin{subfigure}{0.25\textwidth}
        \centering
        $\Qcircuit @C=4em @R=1.8em {
            \lstick{1} & \qw & \qw & \\
            \lstick{2} & \control \qw \ar @{-} [d] ^{\raisebox{0.6ex}{$R_{\rm ZZ}$}} \qwx[1] & \qw & \\
            \lstick{3} & \control \qw & \qw & \\
            \lstick{4} & \qw & \qw & 
        }$
        $\vspace{0.15cm}$
        \caption{}
    \end{subfigure}
    \begin{subfigure}{0.2\textwidth}
        \centering
        $\Qcircuit @C=4em @R=2em {
            \lstick{1} & \control \qw \ar @{-} [d] ^{\raisebox{0.6ex}{$R_{\rm ZZ}$}} \qwx[1] & \qw & \\
            \lstick{2} & \control \qw & \qw & \\
            \lstick{3} & \qw & \qw & \\
            \lstick{4} & \qw & \qw & 
        }$
        $\vspace{0.15cm}$
        \caption{}
    \end{subfigure}
    \begin{subfigure}{0.4\textwidth}
        \centering
        $\Qcircuit @C=4em @R=1.8em {
            \lstick{j} & \control \qw \ar @{-} [d] ^{\raisebox{0.6ex}{$R_{\rm ZZ}$}}  \qwx[3] & \qw \\
            \lstick{k} & \qw & \qw \\
            \lstick{n-1} & \qw & \qw \\
            \lstick{n} & \control \qw & \qw
        }$
        \caption{}
    \end{subfigure}
    \caption{Circuits for the logical Phase gate in the $[[4,2,2]]$ and $[[n,n-2,2]]$ QEDCs. Circuits (a) and (b) show the Phase gate in the $[[4,2,2]]$ QEDC on logical qubit 1 and logical qubit 2, respectively. Circuit (c) shows the logical Phase gate on the $j$th logical qubit in the $[[n,n-2,2]]$ code for $n>4$.}
    \label{fig:Phase}
\end{figure*}

\begin{figure*}[htbp]
    \centering
    \begin{subfigure}{0.43\textwidth}
        \centering
        $\Qcircuit @C=3em @R=1.8em {
            \lstick{1} & \qw & \control \qw \ar @{-} [d] ^{\raisebox{0.6ex}{$R_{\rm XX}$}} \qwx[1] & \qw & \qw & \\
            \lstick{2} & \control \qw \ar @{-} [d] ^{\raisebox{0.6ex}{$R_{\rm ZZ}$}} \qwx[1] & \control \qw & \control \qw \ar @{-} [d] ^{\raisebox{0.6ex}{$R_{\rm ZZ}$}} \qwx[1] & \qw & \\
            \lstick{3} & \control \qw & \qw  & \control \qw & \qw & \\
            \lstick{4} & \qw & \qw & \qw & \qw & 
        }$
        \caption{}
        \label{Hadamard1}
    \end{subfigure}
    \begin{subfigure}{0.43\textwidth}
        \centering
        $\Qcircuit @C=3em @R=1.8em {
            \lstick{1} & \control \qw \ar @{-} [d] ^{\raisebox{0.6ex}{$R_{\rm ZZ}$}} \qwx[1] & \qw  & \control \qw \ar @{-} [d] ^{\raisebox{0.6ex}{$R_{\rm ZZ}$}} \qwx[1] & \qw \\
            \lstick{2} & \control \qw & \control \qw \ar @{-} [d] ^{\raisebox{0.6ex}{$R_{\rm XX}$}} \qwx[1] & \control \qw & \qw \\
            \lstick{3} & \qw & \control \qw & \qw & \qw \\
            \lstick{4} & \qw & \qw & \qw & \qw
        }$
        \caption{}
    \end{subfigure}
    $\vspace{0.5cm}$
    \begin{subfigure}{0.43\textwidth}
        \centering
        $\Qcircuit @C=3em @R=1.8em {
            \lstick{j} & \control \qw \ar @{-} [d] ^{\raisebox{0.6ex}{$R_{\rm ZZ}$}} \qwx[3] & \control \qw \ar @{-} [d] ^{\raisebox{0.6ex}{$R_{\rm XX}$}} \qwx[2] & \control \qw \ar @{-} [d] ^{\raisebox{0.6ex}{$R_{\rm ZZ}$}} \qwx[3] & \qw \\
            \lstick{k} & \qw & \qw & \qw & \qw \\
            \lstick{n-1} & \qw & \control \qw & \qw & \qw \\
            \lstick{n} & \control \qw & \qw & \control \qw & \qw
        }$
        \caption{}
        \label{Hadamardj}
    \end{subfigure}
    \caption{Circuits for the logical Hadamard gate in the $[[4,2,2]]$ and $[[n,n-2,2]]$ QEDCs. Circuits (a) and (b) for the $[[4,2,2]]$ code show the Hadamard gate on logical qubit 1 and logical qubit 2, respectively. Circuit (c) shows the Hadamard gate on the $j$th logical qubit in the $[[n,n-2,2]]$ code for $n>4$.}
    \label{fig:Hadamard}
\end{figure*}

\subsection{Reconciling phases}

We now need to reintroduce phase into our analysis. We can keep track of phases in the binary symplectic representation by defining an additional vector $\underline{\phi}$, which is 2n-dimensional. This vector's entries are $1$,$i$,$-1$, or $-i$. A Clifford operator can multiply these phases independently by $\pm 1$, where each phase is associated with one generator. This is easy to see for the canonical generators $\{X_j\}$ and $\{Z_j\}$, where the subscript $j$ indicates the qubit acted on. We can multiply their phases by $\pm 1$ if we apply a suitable Pauli operator:
\[
Z_j X_j Z_j = -X_j , \quad X_j Z_j X_j = -Z_j ,
\]
\[
Y_j X_j Y_j = -X_j , \quad Y_j Z_j Y_j = -Z_j ,
\] 
since they anticommute. It can be shown that applying the operator $R_{\rm ZZ}$ is equivalent to the symplectic representation of the $R_{\rm ZZ}$ gate without any phase errors. The operator $R_{\rm XX}$ is also exactly equivalent to the symplectic representation of the $R_{\rm XX}$ gate. However, phase errors could still occur in the circuits for the encoded Hadamard, CNOT and Phase gates, since applying an $R_{\rm XX}$ or $R_{\rm ZZ}$ gate twice leaves one of the starting operators with a phase of $-1$. For example, applying the $R_{\rm ZZ}$ gate twice to the operator $XI$ returns $-XI$. We will now determine what phase correction is needed, if any, after an encoded Hadamard, CNOT, and Phase gate made up of these basic gates.

As we show in Appendix~\ref{appendix:nonWeaklyFaultTolerantCircuitWalkthrough}, our construction for the logical Hadamard gate on the $j$th logical qubit, shown in Fig.~\ref{fig:Hadamard}, introduces a phase error of $-1$ in the logical operators. We can fix this in the $[[4,2,2]]$ code by applying the Pauli operators $XXII$ and $IZZI$ after applying a Hadamard gate on logical qubit $1$. For a Hadamard gate on logical qubit $2$, the corrections are $IXXI$ and $ZZII$. This is only true for the special version of the $[[4,2,2]]$ QEDC we used in our paper to achieve a CNOT with just a swap. The logical operators of the code are defined in Eq.~(\ref{eq:422logops}). We can fix this error in the $[[n,n-2,2]]$ QEDC by applying the Pauli operators $Z_j Z_n$ and $X_j X_{n-1}$. Since these are transversal operations they are intrinsically fault-tolerant.

The logical CNOT (shown in Fig.~\ref{fig:CNOT}) is slightly more complicated, since the construction for it differs between the $[[4,2,2]]$ QEDC and the general $[[n,n-2,2]]$ code with $n > 4$. We can use a SWAP to implement a logical CNOT in the $[[4,2,2]]$ code, so there are no phase errors. In the latter case for $n > 4$, one can show through the same procedure outlined in Appendix~\ref{appendix:nonWeaklyFaultTolerantCircuitWalkthrough} that the encoded CNOT still does not produce any phase errors. Finally, our construction for the logical Phase gate (shown in Fig.~\ref{fig:Phase}) introduces no phase errors as well. This result follows directly from the action of the $R_{\rm ZZ}$ gate on the $XI$ operator in Eq.~\ref{eq:phaseTransformations}.

By following our prior constructions with these phase corrections, we can implement the entire encoded Clifford group without phase errors. Moreover, if we choose we can propagate the phase errors forward from an entire Clifford circuit and correct them all at once. Any Pauli error on a single-qubit gate is detectable. As a result, this kind of transversal phase correction is weakly fault-tolerant.

\subsection{Weakly fault-tolerant construction}
\label{sec:weaklyFaultTolerant}

We now have a set of gates that are sufficient to implement any encoded Clifford operation in the $[[n,n-2,2]]$ QEDC. Our next objective is to make them weakly fault-tolerant. It is immediately clear that the SWAP gate is weakly fault-tolerant, since we are assuming that the physical qubits can just be relabeled. The $R_{\rm ZZ}$ and $R_{\rm XX}$ gates, on the other hand, can produce weight-2 errors that commute with the stabilizer generators of the QEDC. A natural question is whether or not fault-tolerant constructions already exist for similar gates that can be used to construct these fixed rotations. Such a construction could automatically be weakly fault-tolerant. The $R_{\rm ZZ}$ gate is closely related to the CZ gate through single-qubit operations. In Ref. \cite{Chao_2018b}, a circuit for a fault-tolerant CZ gate is proposed. Can we construct a weakly fault-tolerant $R_{\rm ZZ}$ using their fault-tolerant CZ gadget? Through common circuit identities presented in Ref.~\cite{nielsen_chuang_2010} and the constructions we offer later in this paper to perform an arbitrary ZZ rotation in Sec.~\ref{sec:Encoded rotations with analog errors}, one could construct an $R_{\rm ZZ}$ gate from two CZ gates, Hadamard gates, and a single qubit $\frac{\pi}{2}$ $R_{\rm Z}$ or $R_{\rm X}$ rotation. However, such a circuit is not fault-tolerant even though the two-qubit gates that make it up are. A single Pauli $Z$ error (in the case of an $R_{\rm Z}$ rotation) or Pauli $X$ error (in the case of an $R_{\rm X}$ rotation) after the rotation gate will evolve into a $ZZ$ error by the end of the circuit. This is an undetectable error in the $[[n,n-2,2]]$ code. Therefore, we cannot construct a weakly fault-tolerant $R_{\rm ZZ}$ gate in this code just with fault-tolerant CZ gates and common circuit identities. We need a new construction to make the $R_{\rm ZZ}$ and $R_{\rm XX}$ gates weakly fault-tolerant.

To make these gates weakly fault-tolerant, we introduce two additional ancillas that are initialized in either the $\ket{\Phi_+}$ or $\ket{++}$ state, where $\ket{\Phi_+}=\sqrt{1/2}(\ket{00}+\ket{11})$ is a Bell state. We can then implement the $R_{\rm ZZ}$ and $R_{\rm XX}$ gates as a sequence of interactions between the two data qubits and the two ancillas. At the end of the circuit, the data qubits are transformed by the desired two-qubit gate ($R_{\rm ZZ}$ or $R_{\rm XX}$), and the ancillas are left in a known quantum state. Any errors produced by a single faulty gate during the circuit will be detectable by measuring the stabilizer generators of either the ancillas or the QEDC at the end of the computation. If an error is detected, the run can be discarded, which allows us to avoid conditional operations in the middle of the circuit, since these are still relatively slow and can induce damaging errors (such as crosstalk) on other data qubits \cite{Hothem2025}.

We identified the weakly fault-tolerant circuits in Figs.~\ref{fig:FTZZ} and \ref{fig:FTXX} using a Mathematica script to search the set of all Clifford circuits up to a limited size for weakly fault-tolerant constructions. These circuits all implement the desired $R_{\rm XX}$ or $R_{\rm ZZ}$ gate on the two data qubits. Most important, any Pauli errors produced by a single faulty gate in these constructions leads to a detectable error at the end. Further details on these circuits are provided in Appendix~\ref{appendix:WeaklyFaultTolerantCircuitWalkthrough}. In the circuit diagrams, an unconnected box with an $R_{\rm X}$ in it refers to the $R_{\rm X}$ operator defined in Eq.~(\ref{eq:RxUyyOperator}).

Interestingly, these circuits cause the ancillas to shift between the $\ket{\Phi_+}$ and $\ket{++}$ states. The $R_{\rm ZZ}$ and $R_{\rm XX}$ gates each have two different constructions as a result. One must therefore keep track of what state the ancillas are left in at each stage in the computation and apply the appropriate form of the encoded gates. In the circuit diagrams, the starting state is on the left and is either $\ket{\Phi_{+}}$ or $\ket{++}$. To detect an error, the ancillas must be measured in the eigenbasis corresponding to the state they should be left in at the end of the computation. By tracing how Pauli errors propagate through these circuits, one can show that a computation composed of these circuits is weakly fault-tolerant. Another way to see this result is by noting that any single fault is detectable at the end of a weakly fault-tolerant gate. Since any subsequent weakly fault-tolerant gates leave the stabilizers of the code unchanged, this error will always remain detectable. Remarkably, this result still holds even if we use the same two ancillas for all of our weakly fault-tolerant $R_{\rm ZZ}$ and $R_{\rm XX}$ gates, though more ancilla qubits can, of course, be used if they are available.

\begin{figure*}[tbp]
    \centering
    \begin{subfigure}{0.8\textwidth}
        \centering
        $\Qcircuit @C=3em @R=1.8em {
            \lstick{\ket{\Phi+}} & \control \qw \ar @{-} [d] ^{\raisebox{0.6ex}{$R_{\rm ZZ}$}} \qwx[3] & \control \qw \ar @{-} [d] ^{\raisebox{0.6ex}{$R_{\rm XX}$}} \qwx[2] & \control \qw \ar @{-} [d] ^{\raisebox{0.6ex}{$R_{\rm ZZ}$}} \qwx[1] & \qw & \control \qw \ar @{-} [d] ^{\raisebox{0.6ex}{$R_{\rm XX}$}} \qwx[2] & \qw & \control \qw \ar @{-} [d] ^{\raisebox{0.6ex}{$R_{\rm ZZ}$}} \qwx[1] & \control \qw \ar @{-} [d] ^{\raisebox{0.6ex}{$R_{\rm ZZ}$}} \qwx[3] & \qw & \qw & \ket{+} \\
            \lstick{} & \qw  & \qw & \control \qw & \control \qw \ar @{-} [d] ^{\raisebox{0.6ex}{$R_{\rm YY}$}} \qwx[1] & \qw & \gate{R_{\rm X}} & \control \qw & \qw & \control \qw \ar @{-} [d] ^{\raisebox{0.6ex}{$R_{\rm YY}$}} \qwx[1] & \qw & \ket{+} \\
            \lstick{\mathrm{data}} & \qw  & \control \qw & \qw & \control \qw & \control \qw & \qw & \qw & \qw & \control \qw & \qw & \mathrm{data}' \\
            \lstick{\mathrm{data}} & \control \qw & \qw & \qw & \qw & \qw & \qw & \qw & \control \qw & \qw & \qw  & \mathrm{data}'
        }$
        \caption{}
        \label{ZZ,Phi+}
    \end{subfigure}
    $\vspace{0.5cm}$
    \begin{subfigure}{0.8\textwidth}
        \centering
        $\Qcircuit @C=3em @R=1.8em {
            \lstick{\ket{+}} & \control \qw \ar @{-} [d] ^{\raisebox{0.6ex}{$R_{\rm ZZ}$}} \qwx[1] & \qw & \control \qw \ar @{-} [d] ^{\raisebox{0.6ex}{$R_{\rm ZZ}$}} \qwx[2] & \control \qw \ar @{-} [d] ^{\raisebox{0.6ex}{$R_{\rm ZZ}$}} \qwx[3] & \control \qw \ar @{-} [d] ^{\raisebox{0.6ex}{$R_{\rm XX}$}} \qwx[2] & \qw & \control \qw \ar @{-} [d] ^{\raisebox{0.6ex}{$R_{\rm XX}$}} \qwx[1] & \control \qw \ar @{-} [d] ^{\raisebox{0.6ex}{$R_{\rm XX}$}} \qwx[2] & \control \qw \ar @{-} [d] ^{\raisebox{0.6ex}{$R_{\rm ZZ}$}} \qwx[3] & \qw & \ket{\Phi+} \\
            \lstick{\ket{+}} & \control \qw & \gate{R_{\rm X}} & \qw & \qw & \qw & \control \qw \ar @{-} [d] ^{\raisebox{0.6ex}{$R_{\rm YY}$}} \qwx[1] & \control \qw & \qw & \qw & \qw & \rstick{} \\
            \lstick{\mathrm{data}} & \qw & \qw & \control \qw & \qw & \control \qw & \control \qw & \qw & \control \qw & \qw & \qw & \mathrm{data}' \\
            \lstick{\mathrm{data}} & \qw & \qw & \qw & \control \qw & \qw & \qw & \qw & \qw & \control \qw & \qw & \mathrm{data}'
        }$
        \caption{}
    \end{subfigure}
    \caption{Weakly fault-tolerant circuits for the $R_{\rm ZZ}$ rotation gate. $R_{\rm YY}$ and $R_{\rm X}$ rotation gates have a fixed angle of $+\frac{\pi}{2}$.}
    \label{fig:FTZZ}
\end{figure*}

\begin{figure*}[tbp]
    \centering
    \begin{subfigure}{0.8\textwidth}
        \centering
        $\Qcircuit @C=3em @R=1.8em {
            \lstick{\ket{\Phi+}} & \control \qw \ar @{-} [d] ^{\raisebox{0.6ex}{$R_{\rm XX}$}} \qwx[3] & \control \qw \ar @{-} [d] ^{\raisebox{0.6ex}{$R_{\rm ZZ}$}} \qwx[2] & \control \qw \ar @{-} [d] ^{\raisebox{0.6ex}{$R_{\rm XX}$}} \qwx[2] & \qw & \control \qw \ar @{-} [d] ^{\raisebox{0.6ex}{$R_{\rm ZZ}$}} \qwx[1] & \control \qw \ar @{-} [d] ^{\raisebox{0.6ex}{$R_{\rm ZZ}$}} \qwx[2] & \qw & \control \qw \ar @{-} [d] ^{\raisebox{0.6ex}{$R_{\rm XX}$}} \qwx[3] & \control \qw \ar @{-} [d] ^{\raisebox{0.6ex}{$R_{\rm ZZ}$}} \qwx[1] & \qw & \ket{+} \\
            \lstick{} & \qw & \qw & \qw & \control \qw \ar @{-} [d] ^{\raisebox{0.6ex}{$R_{\rm YY}$}} \qwx[1] & \control \qw & \qw & \gate{R_{\rm X}} & \qw & \control \qw & \qw & \ket{+} \\
            \lstick{\mathrm{data}} & \qw & \control \qw & \control \qw & \control \qw & \qw & \control \qw & \qw & \qw & \qw & \qw & \mathrm{data}' \\
            \lstick{\mathrm{data}} & \control \qw & \qw & \qw & \qw & \qw & \qw & \qw & \control \qw & \qw & \qw & \mathrm{data}'
        }$
        \caption{}
    \end{subfigure}
    $\vspace{0.5cm}$
    \begin{subfigure}{0.8\textwidth}
        \centering
        $\Qcircuit @C=3em @R=1.8em {
            \lstick{\ket{+}} & \control \qw \ar @{-} [d] ^{\raisebox{0.6ex}{$R_{\rm ZZ}$}} \qwx[1] & \control \qw \ar @{-} [d] ^{\raisebox{0.6ex}{$R_{\rm XX}$}} \qwx[2] & \qw & \control \qw \ar @{-} [d] ^{\raisebox{0.6ex}{$R_{\rm XX}$}} \qwx[3] & \control \qw \ar @{-} [d] ^{\raisebox{0.6ex}{$R_{\rm ZZ}$}} \qwx[2] & \qw & \control \qw \ar @{-} [d] ^{\raisebox{0.6ex}{$R_{\rm ZZ}$}} \qwx[1] & \control \qw \ar @{-} [d] ^{\raisebox{0.6ex}{$R_{\rm ZZ}$}} \qwx[2] & \control \qw \ar @{-} [d] ^{\raisebox{0.6ex}{$R_{\rm XX}$}} \qwx[3] & \qw & \ket{\Phi+} \\
            \lstick{\ket{+}} & \control \qw & \qw & \gate{R_{\rm X}} & \qw & \qw & \control \qw \ar @{-} [d] ^{\raisebox{0.6ex}{$R_{\rm YY}$}} \qwx[1] & \control \qw & \qw & \qw & \qw & \rstick{} \\
            \lstick{\mathrm{data}} & \qw & \control \qw & \qw & \qw & \control \qw & \control \qw & \qw & \control \qw & \qw & \qw & \mathrm{data}' \\
            \lstick{\mathrm{data}} & \qw & \qw & \qw & \control \qw & \qw & \qw & \qw & \qw & \control \qw & \qw & \mathrm{data}'
        }$
        \caption{}
    \end{subfigure}
    \caption{Weakly fault-tolerant circuits for the $R_{\rm XX}$ rotation gate.}
    \label{fig:FTXX}
\end{figure*}

These weakly fault-tolerant circuits were constructed using the binary symplectic representation, so we must separately analyze how they affect the phases of the stabilizer generators and logical operators. Table~\ref{tab:phaseCorrection} shows the recovery operation for each of the 4 gate constructions. It is important to note that
\begin{equation}
R_{\rm X} = \frac{1}{\sqrt{2}}(I - iX) ,\ \ \ \ 
R_{\rm YY} = \frac{1}{\sqrt{2}}(I - iYY) .
\label{eq:RxUyyOperator}
\end{equation}
As was the case with Figs.~\ref{fig:CNOT}--\ref{fig:Hadamard}, the gates depicted involve the unitary versions of the $R_{\rm XX}$, $R_{\rm ZZ}$, $R_{\rm X}$, and $R_{\rm YY}$ gates given in Eqs.~(\ref{ZZ Unitary}), (\ref{XX Unitary}), and (\ref{eq:RxUyyOperator}). Our phase correction operations use only single-qubit gates. Since any single-qubit error is detectable, they are clearly weakly fault-tolerant. Combining these recovery operations with our original constructions completes our circuits for $R_{\rm XX}$ and $R_{\rm ZZ}$. Since the SWAP, $R_{\rm ZZ}$ and $R_{\rm XX}$ gates are sufficient to generate the entire encoded Clifford group in the QEDC, we can carry out any encoded Clifford gate in a weakly fault-tolerant manner.

\begin{table*}[t]
    $\begin{tabular}{ |c|c|c| }
        \hline
         & $R_{\rm ZZ}$ rotation & $R_{\rm XX}$ rotation \\
        \hline
        $\ket{\Phi{+}}$ & ZIZI & IZYI \\
        \hline
        $\ket{++}$ & XIXI & YIYI \\
        \hline
    \end{tabular}$
    \caption{Pauli recovery operations to apply after each weakly fault-tolerant $R_{\rm XX}$ or $R_{\rm ZZ}$ gate. The row is determined by the state the ancilla qubits are in directly before the gate, and the column by which rotation gate we are implementing. Each recovery operation is a 4-qubit operator and can be implemented with 4 single-qubit Pauli gates.}
    \label{tab:phaseCorrection}
\end{table*}

\section{Non-Clifford operations}

To achieve universal quantum computation, it is sufficient to be able to implement any Clifford operation and have one gate outside the Clifford group \cite{nebe_rains_sloane_2006}. The same principle applies to logical operations, so a fault-tolerant non-Clifford gate is required. Unfortunately, this is usually quite difficult and often requires relatively costly protocols, like magic state distillation, to achieve full fault tolerance. Our goal is to avoid this by only enforcing weak fault tolerance and allowing certain analog errors. Since any non-Clifford gate will allow universality, our choice is the Pauli $Z$ rotation gate $R_Z(\theta)$. As we will see, this gate has a straightforward implementation in the $[[n,n-2,2]]$ QEDC. It is also quite commonly used in algorithms for quantum simulation, which are the most common applications of near-term quantum computers. The Pauli $Z$ rotation gate is the operator
\begin{equation}
R_Z(\theta) = \cos\left( {\theta}/{2} \right) I - i \sin\left( {\theta}/{2} \right) Z ,
\end{equation}
for some angle $\theta$. For our model of analog errors, we will assume a faulty physical $Z$ rotation gate applies a rotation by $\theta + \delta \theta$, where $\delta \theta$ is a random variable with 
\[
\mathbb{E} [\delta \theta] = 0 , \quad \quad \mathbb{E}[\delta \theta^2] = \sigma^2 ,
\]
where $\sigma^2$ is the variance. If $\sigma^2 \ll 1$, then by averaging over $\delta \theta$ one can show that a faulty $R_Z(\theta)$ gate is equivalent up to second order to 
\begin{equation}
\begin{aligned}
\ket{\psi} \bra{\psi} &\rightarrow (1-p) R_Z(\theta) \ket{\psi} \bra{\psi} R_Z^\dagger(\theta) \\
&+ p Z R_Z(\theta) \ket{\psi} \bra{\psi}  R_Z^\dagger(\theta) Z ,
\label{eq:analogError}
\end{aligned}
\end{equation}
where $\ket{\psi}$ is an arbitrary qubit state and $p \approx \sigma^2/4$. So analog errors are essentially equivalent to applying a Pauli $Z$ error with some probability after a correct rotation $R_Z(\theta)$. Such errors will not be detectable in the following encoded circuits, but all other single faults will be.

\subsection{Encoded rotations with analog errors}
\label{sec:Encoded rotations with analog errors}

We will now see how to implement a logical Pauli $Z$ rotation in the $[[n,n-2,2]]$ QEDC. It is straightforward to implement a logical $R_Z(\theta)$ gate in this QEDC, if we don't worry about errors. Simply take the logical qubit to be rotated out of the code (by applying a CNOT gate between the data qubit and the parity check qubit), apply a physical $R_Z(\theta)$ gate to the data qubit, and reinsert it back into the code with another CNOT. As one would expect, this removes any protection the logical qubit had from errors during the gate.

We can improve this procedure by adding a few extra gates and an additional ancilla, as shown in Fig.~\ref{fig:logicalRz}. This circuit implements a logical $R_Z(\theta)$ multiplied by the stabilizer generator $Z$ on the ancilla, which is left in the state $\ket0$ at the end of the circuit. We can apply Pauli errors after each of the gates and propagate them through the circuit to see whether they can be detected at the end. All single faults are detectable with three exceptions: a $Z$ error before the $R_Z(\theta)$ gate; a $Z$ error after the $R_Z(\theta)$ gate; or a $ZZ$ error after the third CNOT gate in Fig.~\ref{fig:logicalRz}. By the argument in Eq.~(\ref{eq:analogError}), these are all equivalent to an analog error in the physical $R_Z(\theta)$ rotation gate. 
Since the circuit has some undetectable errors, it is not weakly fault-tolerant in the same sense that the earlier Clifford circuits were. However, this is unavoidable without much costlier non-Clifford constructions that are unlikely to be possible in near-term quantum processors. We consider one such construction below, but others (such as magic state distillation) are also possible.

\begin{figure}[htbp]
    $\Qcircuit @C=1em @R=.7em {
        \lstick{j} & \qw & \targ & \gate{Rz(\theta)} & \targ & \qw & \qw \\
        \lstick{n} & \ctrl{1} & \qw & \qw & \qw & \ctrl{1} & \qw \\
        \lstick{\ket{0}} & \targ & \ctrl{-2} & \qw & \ctrl{-2} & \targ & \qw 
    }$
    \caption{Logical $R_Z(\theta)$ rotation gate on the $j$th logical qubit. This circuit is subject to analog errors: that is, $Z$ errors after the rotation gate are not detectable, but all other single-gate errors are.}
    \label{fig:logicalRz}
\end{figure}

One intuitive way to understand the undetectability of such analog errors is to observe that a $R_Z(\theta)$ followed by a Pauli $Z$ gate is also a valid rotation; or more generally, there is no difference in the circuit between a rotation $R_Z(\theta)$ and a rotation $R_Z(\theta+\delta\theta)$. If $\theta$ is arbitrary we cannot expect to detect errors that are equivalent to simply rotating by the wrong angle. So this logical $R_Z(\theta)$ gate is weakly fault-tolerant up to an imprecision in the physical rotation gate.

\subsection{Probabilistic rotation gates with resource states}

For small quantum computations, the effects of analog errors may be quite tolerable; other errors are more damaging. Over a longer computation such errors can accumulate and derail the calculation. Are there methods, in principle, that could reduce the effects of such errors? We now show that there are, but they demand capabilities beyond the simplest version of weakly fault-tolerant quantum computation that we have considered so far: first, the ability to carry out conditional operations, and second (perhaps) another weakly fault-tolerant non-Clifford gate.

We can improve the logical $R_Z(\theta)$ circuit by introducing a protocol to ensure that the physical rotation gate is not faulty. Up to this point, we have only considered deterministic constructions for logical gates. We can improve the physical rotation gate in our circuit by introducing resource states, an idea that also underlies magic states \cite{Bravyi_2005}. We will make use of a simple repeat-until-success circuit, where each repetition succeeds with probability $1/2$ \cite{Cody_Jones_2012}. Suppose that we can prepare the quantum state
\begin{equation}
\ket{\phi_{\theta}} = \frac{1}{\sqrt{2}} \left(e^{i\theta/2} \ket{0} + e^{-i\theta/2} \ket{1} \right) ,
\label{eq:rotatedResourceState}
\end{equation}
and add it in as an extra qubit in our system. The circuit of Fig.~\ref{fig:resourceState} uses this state to implement a rotation on a physical qubit in our code, while also consuming the resource state. The measurement result tells us whether we rotated by $+\theta$ or $-\theta$. Each occurs with probability $\frac{1}{2}$. If we rotated by the incorrect angle, we can repeatedly apply the circuit with corrections to the angle until we measure that the proper rotation occurred. Unfortunately, it appears that circuits with a higher probability of the correct rotation than $1/2$ are not possible.

\begin{figure}[htbp]
    $\Qcircuit @C=1em @R=.7em {
        \lstick{\ket{\psi}} & \ctrl{1} & \qw & \qw & \quad R_Z(\pm{\theta}) \ket{\psi} \\
        \lstick{\ket{\phi_\theta}} & \targ & \qw & \measureD{Z}
    }$
    \caption{A rotation gate that rotates a quantum state by an angle $\theta$. In this case, $\ket{\psi} = \alpha \ket{0} + \beta \ket{1}$ and $\ket{\phi_\theta}$ is given in Eq.~(\ref{eq:rotatedResourceState}).}
    \label{fig:resourceState}
\end{figure}

The circuit in Fig.~\ref{fig:resourceState} uses only Clifford gates (the CNOT) and Pauli measurements, so an encoded version of this circuit can be done weakly fault-tolerantly in the $[[n,n-2,2]]$ QEDC. Of course, this construction does not really solve the problem of analog errors; it merely changes the difficulty of implementing a logical rotation gate into the difficulty of preparing the resource state. Such a state can be prepared using a rotation gate, but of course it would still have analog errors in the preparation process. Magic states get around this difficulty using distillation, but this generally works only for certain specific resource states for specific rotation angles. A similar approach that works more broadly \cite{Kakkar_2022} is to verify the states by {\it symmetrization}.

\subsection{Resource state symmetrization}
\label{sec:Resource state symmetrization}

The symmetrization idea is conceptually simple. Suppose we attempt to prepare $N$ qubits in the state $\phit$. If our preparation circuit is noisy, then the states we actually prepare may have some components of the orthogonal state $\phitbar$, and may be mixed states. We can reduce the components of undesired states by measuring whether or not the entire collection of $N$ qubits is in the {\it completely symmetric subspace}---that is, the space of all states that are $+1$ eigenstates of all permutations. If this measurement succeeds---which it will with high probability if the noise is low---then the resulting state will be close to $N$ copies of the desired state $\phit$.

Let's see how this works. The states $\{\phit,\phitbar\}$ form a basis for a single qubit. We can expand the Hilbert space of $N$ qubits into subspaces
\begin{equation}
\mathcal{H} = \mathcal{H}_0 \oplus \mathcal{H}_1 \oplus \mathcal{H}_2 \oplus \cdots ,
\end{equation}
where $\mathcal{H}_j$ is the space spanned by all states in which $j$ qubits are in the state $\phitbar$ and $N-j$ qubits are in the state $\phit$. This subspace's dimension is given by the binomial coefficient $C(N,j)$.

Permutations of the qubits do not change how many qubits are in the state $\phit$ and how many in the state $\phitbar$, so these subspaces are invariant under permutations. Moreover, within each subspace $\mathcal{H}_j$ there is exactly one state that is a $+1$ eigenstate of all permutations: the symmetric superposition of all product states with $N-j$ qubits in the state $\phit$ and $j$ qubits in the state $\phitbar$. We will denote these completely symmetric states as $\ket{\Phi^N_{+,j}}$. For example, for $N=3$ and $j=1$ it is this state:
\begin{equation}
\begin{aligned}
\ket{\Phi^3_{+,1}} &= \frac{1}{\sqrt{N}} \bigl(
\phitbar \otimes \phit \otimes \phit
+ \phit \otimes \phitbar \otimes \phit \\
&+ \phit \otimes \phit \otimes \phitbar \bigr) .
\end{aligned}
\end{equation}
The states $\{\ket{\Phi^N_{+,0}}, \ket{\Phi^N_{+,1}}, \ldots, \ket{\Phi^N_{+,N}}$ span the completely symmetric subspace.

To see how symmetrization helps, consider a simple model of preparation errors for resource states. Instead of preparing the correct state $\phit$, we prepare the mixture $(1-p)\phit\braphit + p\phitbar\braphitbar$, where $p < 1/N$ is an error probability. Preparing $N$ qubits gives us the state
\begin{equation}
\begin{aligned}
& (1-p)^N \left(\phit\braphit\right)^{\otimes N}
+ p(1-p)^{N-1} \\
& \times \left( \phitbar\braphitbar \otimes \left(\phit\braphit\right)^{\otimes N-1} + {\rm permutations}\right) \\
& + O(p^2) .
\label{eq:noisyResourceStates}
\end{aligned}
\end{equation}
Now we measure whether the state is in the completely symmetric subspace. If the measurement outcome is positive, then the first term in Eq.~(\ref{eq:noisyResourceStates}) is unchanged; each of the $N$ states in the $j=1$ subspace is projected onto $(1/N)\ket{\Phi^N_{+,1}}\bra{\Phi^N_{+,1}}$; and so forth. So the infidelity of the state with $N$ perfect copies of $\phit$ is reduced by roughly $1/N$. The fidelity of each individual resource state goes from $1-p$ to approximately $1-p/N$. A similar conclusion will apply to other error models, so long as the error rate is low and the noise does not preserve the symmetry of the joint state. One example of such a global symmetry-preserving noise model is a coherent error that causes the same over- or under-rotation on all $N$ copies of the state. In this case, the joint state would be symmetric, and as a result the SWAP measurements would not detect the error.

How can such a measurement be done? This is easiest to see for $N=2$. In this case, the only nontrivial permutation is the SWAP. We can measure its eigenvalue with the circuit in Fig.~\ref{fig:symmetryMeas}. For $N>2$ it is a bit more complicated, since permutations are unitary but not necessarily Hermitian, so they are not observables in general. The complete set of permutations grows like $N!$ as well, which suggests that a large number of measurements might be needed. However, all permutations can be generated from a set of $N-1$ pairwise SWAPS. If a state is a simultaneous $+1$ eigenstate of all $N-1$ pairwise SWAPs then it is completely symmetric. So this measurement can be done with $N-1$ copies of the circuit in Fig.~\ref{fig:symmetryMeas}. It is important to note that the pairwise SWAPS do not all commute, but in spite of this they do have simultaneous $+1$ eigenstates.

\begin{figure}[htbp]
    $\Qcircuit @C=1em @R=.7em {
        \lstick{1} & \ctrl{1} & \gate{H} & \ctrl{1} & \gate{H} & \ctrl{1} & \qw \\
        \lstick{2} & \targ & \qw & \ctrl{1} & \qw & \targ & \qw \\
        \lstick{\ket{0}} & \qw & \qw & \targ & \qw & \qw & \measureD{Z} 
    }$
    \caption{Measurement of the eigenvalues $\pm1$ for the SWAP. This determines if the state of qubits 1 and 2 is completely symmetric for the case $N=2$.}
    \label{fig:symmetryMeas}
\end{figure}

\begin{figure*}[htbp]
    $\Qcircuit @C=1em @R=1em {
        \lstick{\ket{0}} & \qw & \qw & \qw & \qw & . & . & . & & \qw & \targ & \qw & \qw \\
        \lstick{\ket{0}} & \qw & \qw & \qw & \qw & . & . & . & & \targ & \qw & \qw & \qw \\
        \lstick{.} & & & & & & &  . & & & & & .\\
        \lstick{.} & & & & & & . & & & & & & . \\
        \lstick{.} & & & & & . & & & & & & & . \\
        \lstick{\ket{0}} & \qw & \qw & \targ & \qw & . &. & . & & \qw & \qw & \qw \qw & \qw \\
        \lstick{\ket{+}} & \ctrl{2} & \qw & \qw & \qw & . &. & . & & \qw & \qw & \ctrl{2} & \qw \\
        \lstick{\ket{0}} & \qw & \targ & \qw & \qw & . &. & . & & \qw & \qw & \qw & \qw \\
        \lstick{\ket{0}} & \targ & \ctrl{-1} & \ctrl{-3} & \qw & . &. & . & & \ctrl{-7} & \ctrl{-8} & \targ & \measureD{Z}
    }$
    \caption{Fault-tolerant circuit for initialization into the $[[n,n-2,2]]$ QEDC starting state (the GHZ state). We assume that our $n-2$ data qubits start in the state $\ket{0}$, and the two check qubits in the states $\ket+$ and $\ket0$; there is also one additional ancilla (the bottom qubit) initialized in state $\ket{0}$. Chao and Reichardt derived an equivalent circuit previously that fault-tolerantly prepares the same logical state \cite{Chao_2018}. This encoding circuit is only weakly fault-tolerant in our protocol, since we measure the flag ancilla only at the end of the computation.}
    \label{fig:initialize}
\end{figure*}

To be useful for fault-tolerant (or weakly fault-tolerant) quantum computation, one would need to use an encoded version of the circuit in Fig.~\ref{fig:symmetryMeas}. Because this circuit includes a Toffoli gate, this is a challenge: a weakly fault-tolerant encoded Toffoli would be required. A fault-tolerant construction for the encoded CCZ gate in the $[[n,n-2,2]]$ code was proposed by Chao and Reichardt \cite{Chao_2018b}, which when conjugated by encoded Hadamard gates on the target qubit is equivalent to an encoded Toffoli gate \cite{Paetznick_2013}. By replacing these encoded Hadamard gates with our weakly fault-tolerant ones, we get a weakly fault-tolerant encoded Toffoli which can be used for the symmetrization procedure. However, the fault-tolerant encoded CCZ proposed in Ref. \cite{Chao_2018b} requires $16$ physical CCZ gates and $50$ two-qubit gates for the encoded CCZ gate. Combined with our constructions for the encoded Hadamard gate, we will require $114$ multi-qubit gates to realize an encoded weakly fault-tolerant Toffoli gate. Since this construction includes many CNOTs and CCZs, it will likely be exceedingly costly to implement on architectures where our scheme is beneficial such as ion traps. Whether a simpler construction exists that only achieves weak fault tolerance rather than full fault tolerance on the CCZ gate is unknown, but it may be possible. This would, of course, open up another avenue to universality since the Toffoli is a non-Clifford gate. However, from a practical point of view it is unclear whether these more complicated circuits make sense for near-term quantum processors, as they demand conditional operations and larger numbers of qubits to hold the resource states. Quantum processors with sufficient resources and capabilities to carry this out may be able to achieve full fault tolerance, which is ultimately needed for scalability.

\section{Initialization and readout}

In the previous sections, we have seen how to construct a set of gates that achieve universal quantum computation. We also showed that these gates are weakly fault-tolerant, if we allow for analog errors in our logical $R_Z(\theta)$ gate. To make this a complete protocol for quantum computation, we also need a procedure to initialize our quantum state in the QEDC at the beginning and to read out our result at the end. We also need to measure the stabilizer generators of the code and the additional ancillas at the end to detect if errors occurred. While this may seem straightforward, due to the simplicity of the QEDC, care is required: we do not want to introduce new errors that could spoil weak fault tolerance. In this section, we will see how to do this in a way that does not allow any single-gate error during initialization or readout to become an undetectable error.

To begin an encoded quantum computation, we must first initialize the qubits into a known quantum state that has the same stabilizer generators as the QEDC. For the $[[n,n-2,2]]$ code, the $n$-qubit GHZ state satisfies this requirement. This is the quantum state $\frac{1}{\sqrt{2}}(\ket{00...0} + \ket{11...1})$, which represents all logical qubits in the state $\ket0$. Most quantum computers begin in a standard starting state, often $\ket{00...0}$. Therefore, we need an encoding unitary on the $n$ qubits to transform this state into the GHZ state in a weakly fault-tolerant manner. The unitary encoding circuit from Fig.~\ref{fig:initialize} does the job for an $n$-qubit initial state using one additional ancilla. A simple analysis using the Pauli error model shows that this circuit is weakly fault-tolerant. It is important to note that the second-to-last qubit, in the state $\ket+$, can be prepared by a single Hadamard gate, which is still weakly fault-tolerant. Some single gate faults are undetectable, but only produce a global phase of $\pm1$, and hence are not errors. Error detection is done by measuring the stabilizer generators of the QEDC, and $Z$ on the single additional $\ket{0}$ ancilla, at the end of the computation. As a result, our implementation of the initialization circuit is only weakly fault-tolerant. One can choose to measure the flag at the end of the state preparation gadget as depicted in Fig.~\ref{fig:initialize}. This mid-circuit measurement makes the initialization circuit fault-tolerant and equivalent to a fault-tolerant circuit by Chao and Reichardt that prepares the same logical state \cite{Chao_2018}.  Any single-qubit initialization error (preparing $\ket1$ instead of $\ket0$) can also be detected. If the starting state differs from $\ket{00...00}$, one must first find an additional weakly fault-tolerant unitary that transforms the starting state into the $\ket{00...00}$ state, followed by the encoding unitary in Fig.~\ref{fig:initialize} to initialize the qubits into the $n$-qubit GHZ state.

\begin{figure*}[htbp]
    $\Qcircuit @C=1em @R=1em {
        \lstick{1} & \qw & \qw & \qw & \qw & . & . & . & & \qw & \targ & \qw & \qw & \measureD{Z} & \multigate{7}{\Sigma} \\
        \lstick{2} & \qw & \qw & \qw & \qw & . & . & . & & \targ & \qw & \qw & \qw & \measureD{Z} & \cghost{\Sigma} \\
        \lstick{.} & & & & & & & . & & & & & & . & \\ 
        \lstick{.} & & & & & & . & & & & & & & . & \\
        \lstick{.} & & & & & . & & & & & & & & . & \\
        \lstick{n-2} & \qw & \qw & \targ & \qw & . &. & . & & \qw & \qw & \qw \qw & \qw & \measureD{Z} & \cghost{\Sigma} \\
        \lstick{n-1} & \ctrl{2} & \qw & \qw & \qw & . &. & . & & \qw & \qw & \ctrl{2} & \qw & \measureD{X} & \\
        \lstick{n} & \qw & \targ & \qw & \qw & . &. & . & & \qw & \qw & \qw & \qw & \measureD{Z} & \cghost{\Sigma} \\
        \lstick{\ket{0}} & \targ & \ctrl{-1} & \ctrl{-3} & \qw & . &. & . & & \ctrl{-7} & \ctrl{-8} & \targ & \qw & \measureD{Z}
    }$
    \caption{Weakly fault-tolerant circuit for readout of the $[[n,n-2,2]]$ QEDC. The $\Sigma$ subcircuit represents a classical parity check decoder that adds the $n-2$ data bits to the final ($n$th) check bit. This circuit measures all $n-2$ of the data qubits in the Z basis as well as the $(n-1)$th check qubit (in the $X$ basis) and the $n$th check qubit (in the $Z$ basis).}
    \label{fig:readout}
\end{figure*}

We now need a way to read out the result of our computation at the end, and measure the stabilizer generators of the QEDC and the additional ancillas we used to detect errors. For the additional ancillas, this is usually straightforward and requires no additional machinery: simply measure projectively in the ancilla's standard basis (usually either $Z$ or $X$). For the pair of ancillas used for the weakly fault-tolerant $R_{\rm ZZ}$ and $R_{\rm XX}$ gates in Figs.~\ref{fig:FTZZ} and \ref{fig:FTXX}, we may need to measure a pair of ancillas in the Bell basis ($\ket{\Phi_+}$ indicating no error).  This can be done weakly fault-tolerantly using one additional ancilla, as shown in Fig.~\ref{fig:ancillaMeas}. It is also important to note that, while our gate gadgets locally satisfy Gottesman's definition of fault tolerance, our weakly fault-tolerant measurement gadgets do not. This is because we do not measure our syndromes redundantly. As a result, a single measurement error can flip the syndrome measurement without detection. In our setting of postselected error detection such an error is not damaging. A syndrome measurement error will either leave the measurement result unchanged or flip the measured eigenvalue. Since we postselect on all stabilizer measurements yielding values of $+1$, the run will be discarded in the case of a measurement result of $-1$. This causes the discard rate to scale linearly in $p$, but that is already the case in our weakly fault-tolerant constructions. A measurement error is much more damaging in an error correction setting, since an erroneous measurement could lead to an incorrect correction that introduces a logical error. The most damaging possible outcome in a postselected error detection protocol is when a measurement error coincides with a previous gate error to incorrectly report a correct computation. However, this is an $O(p^2)$ event and does not violate weak fault tolerance. A procedure showing how to measure $XX$ and $ZZ$ non-destructively is given in the appendix of \cite{Chao_2018}. Their measurement procedure can be adapted to non-destructively measure a pair of ancillas in the Bell basis.

\begin{figure}[htbp]
    $\Qcircuit @C=1em @R=.7em {
        \lstick{1} & \ctrl{2} & \qw & \ctrl{2} & \measureD{X} \\
        \lstick{2} & \qw & \targ & \qw & \measureD{Z} \\
        \lstick{\ket{0}} & \targ & \ctrl{-1} & \targ & \measureD{Z} 
    }$
    \caption{Weakly fault-tolerant Bell basis measurement of qubits 1 and 2.}
    \label{fig:ancillaMeas}
\end{figure}


To read out the data (in the $Z$ basis) and measure the stabilizer generators, we apply the decoding circuit in Fig.~\ref{fig:readout}. This is essentially the inverse of the encoding unitary, using a single ancilla that can be the same as the one used in state preparation. A straightforward error analysis shows that this circuit is weakly fault-tolerant. The subcircuit labeled $\Sigma$ is a classical decoding step for the classical parity-check code: the first $n-2$ bits hold the readout values of the circuit, and the overall parity of those $n-1$ bits should be even if there are no errors. Note that weak fault tolerance is preserved even if we use the same additional ancilla for initialization and readout. This means that we can just measure it at the end of the computation to catch errors.

\section{Resource use and error rate}

\subsection{Resource use}

Having outlined the protocol for weakly fault-tolerant quantum computation, we can now analyze its resource consumption rate. As mentioned earlier, the same ancilla is used for initialization and readout. Our constructions for the $R_{\rm XX}$ and $R_{\rm ZZ}$ rotation gates require two additional ancillas. This isn't costly, since we can reuse the same two ancillas for all of our two-qubit rotation gates and maintain weak fault tolerance. We also require one final additional ancilla for the logical $R_Z(\theta)$ gate, since weak fault tolerance is maintained (up to analog errors) even if we reuse the ancilla.  Putting all of this together gives us a total consumption of $4$ ancillas for an entire quantum computation using our QEDC. This is beneficial for current NISQ machines, since the protocol allows for protection against errors in our logical Clifford circuits up to first order and some error suppression in non-Clifford logical rotations. If we instead use the construction for a probabilistic rotation gate using resource states, then our protocol does begin to consume a sizeable number of $\phit$ quantum states. We would on average consume roughly two of these states for every physical rotation gate. Despite this, the protocol is relatively resource efficient due to the ability to reuse ancillas throughout a computation and the high rate of the code.

In addition to requiring only a small number of ancillas, the code rate approaches 1 in the limit of a large number of encoded qubits. We are using an $[[n,n-2,2]]$ QEDC; the extra check qubits we use to encode our physical qubits are fixed at two, which becomes negligible as we begin to encode more qubits. The additional four ancillas (if we allow for analog errors) is also a fixed number. This gives us a code rate $(n-6)/n$ that approaches 1 as the number of encoded qubits becomes large. This is beneficial for current NISQ machines, since we can use almost all of the physical qubits to represent actual data qubits.

The number of gates required will depend on the original (ideal) circuit. One can decompose the ideal circuit into Clifford gates and $Z$ rotations and then express the encoded Clifford gates using SWAP, $R_{\rm XX}$, and $R_{\rm ZZ}$ gates. The weakly fault-tolerant version of the SWAP gate is still just a single SWAP, but the weakly fault-tolerant $R_{\rm XX}$ and $R_{\rm ZZ}$ gate constructions use nine gates each. This does magnify the depth of a quantum circuit, but it is manageable for short calculations that result in relatively small quantum circuits.
\begin{figure*}[htbp]
    \centering
    \begin{subfigure}{0.48\textwidth}
        \centering
        \includegraphics[scale=0.5]{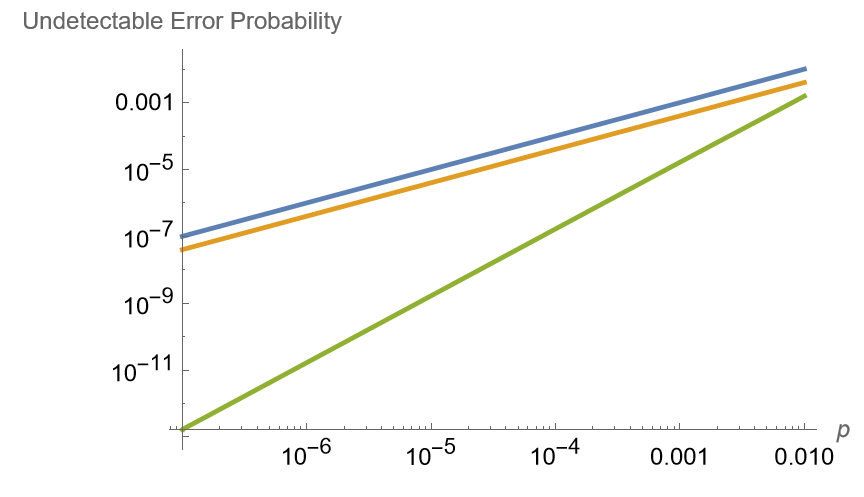}
        \caption{}
    \end{subfigure}
    \begin{subfigure}{0.48\textwidth}
        \centering
        \includegraphics[scale=0.5]{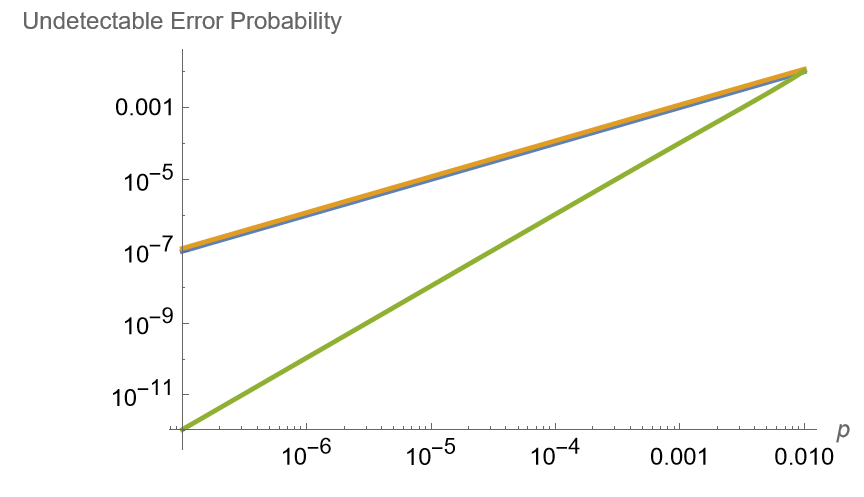}
        \caption{}
    \end{subfigure}
    \caption{Log-log plot of the undetectable error probability for the physical (blue), encoded (orange), and weakly fault-tolerant encoded (green) circuits as a function of the physical error probability $p$. The above plots are the probabilities of an undetectable error for (a) the Hadamard gate and (b) the CNOT gate. The physical and encoded undetectable error probability lines overlap in plot (b). $O(p^2)$ scaling in the physical error rate $p$ is also a hallmark of fully fault-tolerant protocols. Our Clifford gate circuits do satisfy the Gottesman definition of a fault-tolerant gate gadget when combined with postselection. However, we call our gate gadgets weakly fault-tolerant since we only measure the flag ancillas at the end of a computation. For a rigorous justification, see Sec.~\ref{sec:QEDC and weak fault tolerance}.}
    \label{fig:HadamardandCNOTErrorProb}
\end{figure*}
    
\begin{figure*}[htbp]
    \centering
    \begin{subfigure}{0.48\textwidth}
        \centering
        \includegraphics[scale=0.45]{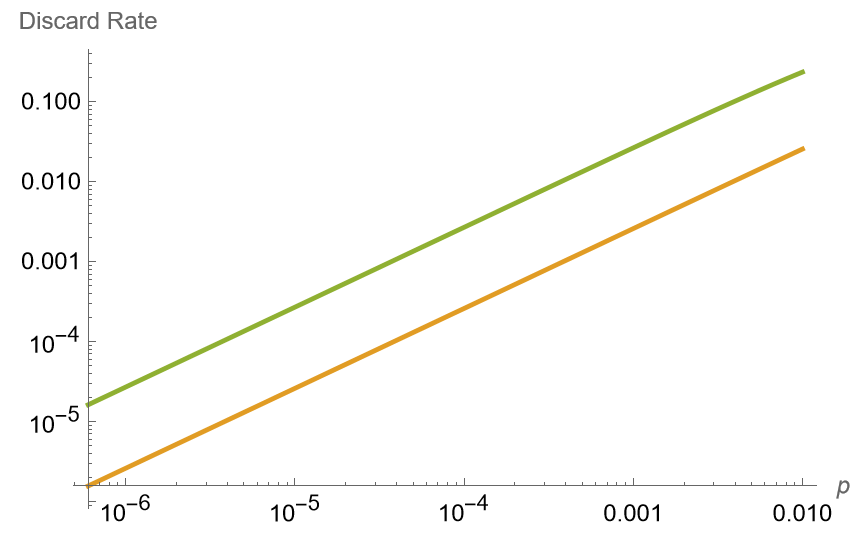}
        \caption{}
    \end{subfigure}
    \begin{subfigure}{0.48\textwidth}
        \centering
        \includegraphics[scale=0.45]{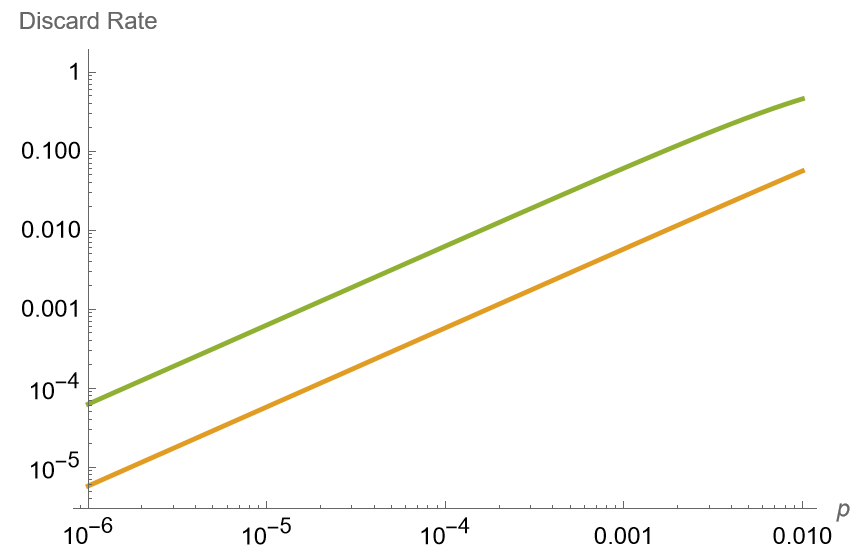}
        \caption{}
    \end{subfigure}
    \caption{Log-log plot of the discard rate for the encoded (orange) and weakly fault-tolerant encoded (green) circuits as a function of the physical error probability $p$. The above plots are the discard rates for (a) the Hadamard gate and (b) the CNOT gate. The discard rate for the physical (blue) circuits is not shown because it is trivially 0.}
    \label{fig:HadamardandCNOTDiscardRate}
\end{figure*}
\subsection{Error and Postselection rate}

Our protocol for quantum computation has relatively low overhead, but we need to analyze its performance as a function of the error rate. For simplicity, we assume a depolarizing error model and that errors on different gates are independent and equally likely. Let us denote the probability of a single gate error as $p$ (the physical error rate of our gates). The main difficulty is determining which errors are undetectable at the end of the circuit. The number of possible errors grows rapidly as the size of the circuit increases. To keep the analysis manageable, we only consider errors up to third order in $p$, and assume that any fault involving errors on four or more different gates is undetectable. From this, one can obtain an upper bound on the probability of an undetectable error for the given circuit.

For the analysis, we used a program (a Mathematica script) to determine what fraction of errors at each order in $p$ are undetectable. We generate the errors, evolve them to the end of the circuit, and determine if they are undetectable. For this analysis, we do not include errors during initialization and readout, but of course they are also important. Plots of this upper bound on the probability of error for the encoded Hadamard and CNOT are shown in Fig.~\ref{fig:HadamardandCNOTErrorProb}. The undetectable-error probability is calculated up to third order. In each case, we consider the probability of an undetectable error for the physical gate, the encoded but not weakly fault-tolerant gate, and the weakly fault-tolerant encoded gate. The third-order approximation converges to the true undetectable error probability as the physical error rate $p$ decreases. In this case, encoded refers to the logical version of the gate in the $[[n,n-2,2]]$ QEDC implemented with the SWAP, $R_{\rm XX}$, and $R_{\rm ZZ}$ gate set. Circuit decompositions of the encoded CNOT and encoded Hadamard in terms of these three gates are given in Figs.~\ref{fig:CNOT} and \ref{fig:Hadamard}, respectively. For a breakdown of how these calculations were performed, see Appendix~\ref{appendix:undetectableErrorCalculation}. The analysis shows that the weakly fault-tolerant constructions for the encoded Hadamard and CNOT gates are a significant improvement over both their unencoded and encoded analogues for physical error probabilities lower than $0.001$. The rate at which the undetectable-error probability decays with the physical error probability is also significantly better for the weakly fault-tolerant gates.

Although the weakly fault-tolerant encoded gates have a lower probability of an undetectable error, it is important to note that this improvement comes at the cost of a higher discard rate. One of the trade-offs made by the fact that our code only detects errors instead of correcting them is that we have to postselect on the outcome from measuring the stabilizer generators of the QEDC and the ancillas. Errors can no longer be corrected because we cannot uniquely determine what error occurred, so erroneous runs must be discarded. Plots of the discard rate are shown in Fig.~\ref{fig:HadamardandCNOTDiscardRate} using the same color scheme for each of the protocols as in Fig.~\ref{fig:HadamardandCNOTErrorProb}. As one would expect, replacing each two-qubit gate with 8 two-qubit gates to achieve weak fault tolerance roughly increases the rejection rate by a factor of 8. The values for the discard rate were obtained by taking the total probability of a detectable error up to third order in $p$ and subtracting this value from 1. Such a scheme is roughly equivalent to always throwing away computational runs where one detects an error. The true discard rate will be different, since we are only considering errors up to third order detectable. However, the accuracy of this approximation increases as the physical error rate decreases. The requirement that we run a circuit many times isn't too costly for smaller circuits that are likely to be implemented on near-term quantum computers, and most such near-term algorithms require multiple runs anyway. The above results show that our scheme for weakly fault-tolerant quantum computation lies in the middle ground that we aimed to fill while we wait for the era of full fault tolerance.

\begin{figure*}[htbp]
    \centering
    \begin{subfigure}{0.48\textwidth}
        \centering
        \includegraphics[scale=0.5]{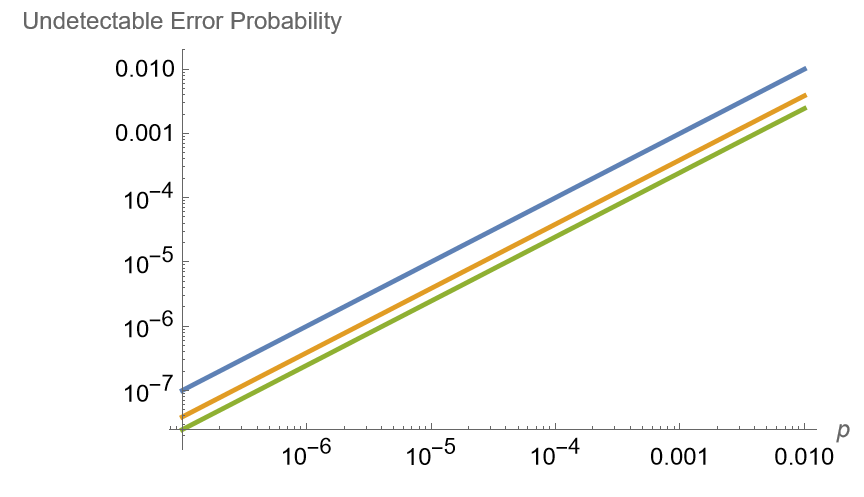}
        \caption{}
    \end{subfigure}
    \begin{subfigure}{0.48\textwidth}
        \centering
        \includegraphics[scale=0.45]{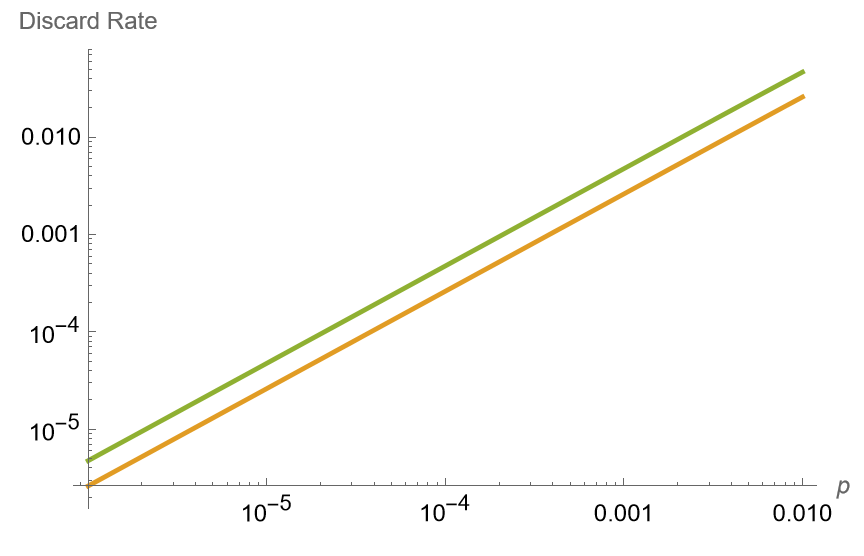}
        \caption{}
    \end{subfigure}
    \caption{Log-log plot of (a) the undetectable error probability and (b) the discard rate for the physical (blue), encoded (orange), and weakly fault-tolerant encoded (green) circuits as a function of the physical error probability $p$. The above plots are for the encoded rotation circuit shown in Fig.~\ref{fig:logicalRz} to implement a non-Clifford gate. }
    \label{fig:Rotations_With_analog_ErrorsPerformanceAnalysis}
\end{figure*}

We will now analyze the performance of the non-Clifford gates. The three protocols for logical rotation gates proposed in this paper are encoded rotations with analog errors, probabilistic rotation gates with resource states, and resource state symmetrization. We will only consider the first two and leave the performance analysis of the symmetrization protocol as potential future work. We begin with encoded rotations with logical errors. Specifically, we are referring to the quantum circuit depicted in Fig.~\ref{fig:logicalRz}. Through the exact same procedure described previously for finding the probability of an undetectable error in the Clifford gates, one can derive the probability of an undetectable error and the discard rate for the given circuit. The analytical results are shown in Fig.~\ref{fig:Rotations_With_analog_ErrorsPerformanceAnalysis}. 

We will now consider our implementation of a logical rotation gate that uses resource states. In order to implement a logical rotation in our QEDC weakly fault-tolerantly with this protocol, we need access to a supply of the logical version of the resource state given in Eq.~(\ref{eq:rotatedResourceState}). If we can satisfy this requirement, then the only remaining source of error lies in the logical CNOT gate in the idealized case of perfect measurement, which is an assumption we have made throughout our analysis. The probability of an undetectable error for this scenario is given in Fig.~\ref{fig:HadamardandCNOTErrorProb} for the three implementations of the CNOT we have considered. For the resource state protocol to make any practical sense, we must have access to mid-circuit measurements. Assuming this is available, the number of logical CNOTS and logical resource states required for the protocol of Fig.~\ref{fig:resourceState} will follow a geometric distribution with a parameter $p=0.5$, which is just equivalent to the number of flips of a fair coin until heads shows up. The discard rate is then just this value subtracted from unity. To extend our previous analysis to this case, one would have to find the numerical distribution of the probability of an undetectable error and the postselection rate. Since detectable errors in different weakly fault-tolerant logical CNOTs can combine to become undetectable errors, such a calculation would require more difficult numerical methods than the simulation methods we have used previously. One possible way to tractably approach this issue is to use Monte Carlo simulations, but we will leave such an analysis for future work. If we assume detectable errors between different weakly fault-tolerant CNOTs cannot become undetectable errors, then the geometric distribution of the probability of an undetectable error and the discard rate can be derived directly from Figs.~\ref{fig:HadamardandCNOTErrorProb} and \ref{fig:HadamardandCNOTDiscardRate} respectively.

We have chosen not to include a numerical analysis for the symmetrization procedure. The main reason for this omission is due to the complexity of the weakly fault-tolerant version of this protocol for even the simplest non-trivial case of size $N=2$. The circuit to measure the eigenvalues of the SWAP is provided in Fig.~\ref{fig:symmetryMeas}, which is necessary to determine if the state is in the completely symmetric subspace. Using our only currently known construction for the weakly fault-tolerant Toffoli gate with a fault-tolerant CCZ gate from Ref.~\cite{Chao_2018b}, the weakly fault-tolerant version of this circuit will contain $274$ multi-qubit gates. Such a circuit would already be exceedingly difficult to analyze with the methods presented in Appendix~\ref{appendix:undetectableErrorCalculation}. To make matters worse, the presence of $16$ non-Clifford CCZ gates will evolve Pauli errors to linear combinations of Pauli errors. As a result, we conclude that such a numerical analysis is infeasible with the simulation methods considered in this paper. More advanced techniques such as Monte Carlo methods are much more suitable for simulating these large quantum circuits that contain non-Clifford gates, but we have not employed this approach in our paper. Since we lack a numerical analysis, we cannot determine the physical error rate for which a weakly fault-tolerant symmetrization procedure makes sense in terms of decreasing the probability of an undetectable error. It is fair to say that a far more efficient implementation of the weakly fault-tolerant Toffoli gate must be found for the procedure to make sense from a practical standpoint. One promising future direction to achieve this is through code-switching. In such a scheme, one would temporarily switch to another code where the Toffoli or CCZ gate is transversal. This is a promising direction for future efforts to make the symmetrization procedure practical for near term quantum computers. Despite the lack of a numerical analysis, it is clear from the mathematical arguments in Sec.~\ref{sec:Resource state symmetrization} that the symmetrization procedure will improve the state fidelity as the number of copies of the state $N$ becomes large.

\begin{figure*}[htbp]
    \centering
    \begin{subfigure}{0.48\textwidth}
        \centering
        \includegraphics[scale=0.5]{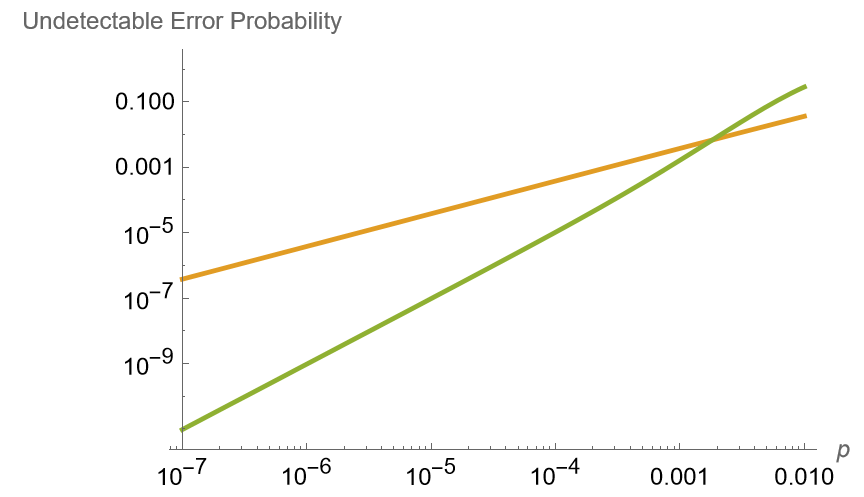}
        \caption{}
    \end{subfigure}
    \begin{subfigure}{0.48\textwidth}
        \centering
        \includegraphics[scale=0.45]{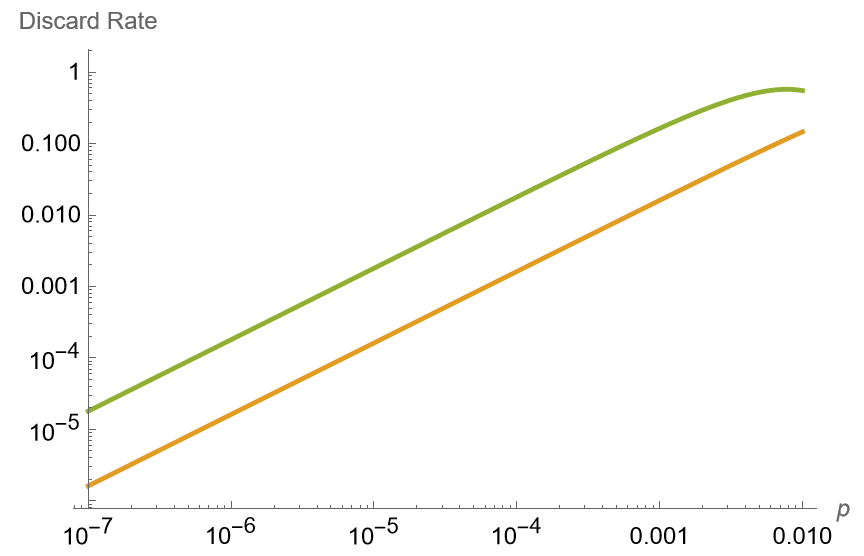}
        \caption{}
    \end{subfigure}
    \caption{Log-log plot of (a) the undetectable error probability and (b) the discard rate for the encoded (orange) and weakly fault-tolerant encoded (green) Bell state preparation mirror circuit shown in Fig.~\ref{fig:Encoded_Bell_State_Mirror_Circuit} as a function of the physical error probability $p$. All values were calculated up to second order in $p$. The encoded version of the circuit involves the gates from Ref. \cite{self2022protecting} with $R_{\rm ZZ}$ and $R_{\rm XX}$ rotation gates having a fixed angle of $+\frac{\pi}{2}$ and $+\frac{3\pi}{2}$, respectively. For a physical error rate near $0.01$, we can see the discard rate start to flatten for our weakly fault-tolerant protocol. This tells us that our weakly fault-tolerant circuit is beginning to be overwhelmed by the combination of the physical error rate and the large number of possible fault locations. The above behavior is expected and occurs for a lower error rate in our protocol as compared to \cite{self2022protecting} due to the circuit magnification caused by our weakly fault-tolerant constructions.}
    \label{fig:Encoded_Bell_State_Mirror_Circuit_Performance_Analysis}
\end{figure*}

For a complete encoded circuit the analysis is more complex, because multiple errors in different parts of the circuit could combine to be undetectable. In the simplest version of the weakly fault-tolerant approach, where the syndromes are checked only at the end of the computation, this is unavoidable. If more ancillas or mid-circuit measurements are available, one could add additional stabilizer checks at extra points during the circuit; this would allow one to detect more errors, but at some cost in increased complexity and overall failure probability.

Finally, we will conclude our analysis by comparing our protocol to the one developed by Self, Benedetti and Amaro \cite{self2022protecting}. Their performance analysis mainly focuses on what they refer to as the ``population survival rate'' and ``discard rate.'' The first term refers to the probability that all of the logical qubits end up in the logical quantum state $\ket{\bar{0}}$ after a mirror circuit. For those who are unfamiliar, a {\it mirror circuit} consists of applying a unitary followed by its inverse. In comparing our approaches, we will take our non-Clifford gate to be the one presented in Fig.~\ref{fig:logicalRz}. The first major difference between these protocols is that ours is designed to achieve weak fault tolerance. As stated earlier, this corresponds to being able to detect any Pauli error produced by any single gate anywhere in an encoded quantum circuit. Our protocol achieves this up to analog errors in the non-Clifford rotation gates. 

From our analysis, it is reasonable to suggest that our weakly fault-tolerant Clifford gates will offer a significant reduction in the probability of an undetectable error in comparison to the gate set contained in \cite{self2022protecting} for a low enough physical error probability. For non-Clifford gates, we expect the probability of an undetectable error in our two protocols to exhibit the same scaling rate in the physical error rate $p$. Nevertheless, our approach will yield a modest reduction in this probability for sufficiently small $p$. Finally, our weakly fault-tolerant initialization and readout circuits are similar to the fault-tolerant ones presented in \cite{self2022protecting}, although we developed them independently. Both achieve the same goal of detecting any error produced by a single faulty gate, so we expect their performance to be similar. It is important to note that while our circuits achieve a similar notion of gadget level fault tolerance, we refer to our constructions as weakly fault-tolerant because all flag and stabilizer measurements are deferred to the end of a computation. Weak fault tolerance is rigorously defined and justified in Sec.~\ref{sec:QEDC and weak fault tolerance}. A fault-tolerant mid-circuit syndrome check circuit is also presented in \cite{self2022protecting}, which we do not include in our protocol to avoid mid-circuit measurements. The large improvement in the error scaling for our weakly fault-tolerant Clifford gates comes at the cost of a significant reduction in the survival probability and postselection rate, as we will show. For the purpose of comparing our two protocols and determining how our weakly fault-tolerant gates affect the number of circuit layers, we must impose some restrictions.

\begin{figure}[htbp]
    $\Qcircuit @C=1em @R=.7em {
        \lstick{\ket{0}} & \gate{H} & \ctrl{1} & \ctrl{1} & \gate{H} & \qw & \ket{0} \\
        \lstick{\ket{0}} & \qw & \targ & \targ & \qw & \qw & \ket{0} \\
    }$
    \caption{Simple mirror circuit that prepares the Bell state $\ket{\Phi+}$ from $\ket{00}$ and then returns the state back to $\ket{00}$.}
    \label{fig:Encoded_Bell_State_Mirror_Circuit}
\end{figure}

For the purposes of this analysis, we will only consider Clifford operations and allow for the equivalence of our native gate sets in this regime. Explicitly, our restriction is equivalent to enforcing that the allowable rotations are multiples of $\frac{\pi}{2}$ in terms of the exponential two-qubit Pauli operators $\exp(-i\theta Z_{i}Z_{j}/2)$ and $\exp(i\theta X_{i}X_{j}/2)$. This second $R_{\rm XX}$ rotation operator can be obtained from the $R_{\rm ZZ}$ rotation operator in combination with other single-qubit rotation gates. As shown earlier, this set of gates is sufficient to implement the logical CNOT. As far as we are aware, it takes $7$ of these two-qubit gates to implement a logical CNOT. In our native gate set, the weakly fault-tolerant CNOT consists of $63$ gates. Looking at Figs.~\ref{fig:CNOT}, \ref{fig:FTXX} and \ref{fig:FTZZ}, one obvious way to minimize the number of layers is to run the single-qubit gates in parallel with one of the two-qubit gates. From this, the weakly fault-tolerant CNOT is equivalent to 56 layers. So each of our weakly fault-tolerant gates will magnify the number of layers of the equivalent gates from \cite{self2022protecting} by about a factor of $8$. We can now offer a comparison of the survival and postselection rates of the two protocols.

To numerically compare our two protocols, we consider a simple quantum circuit that prepares an encoded $\ket{\Phi+}$ state from the logical $\ket{00}$ starting state. We then apply the reverse of this circuit to return the state to the logical $\ket{00}$ state. The unencoded version of the circuit is depicted in Fig.~\ref{fig:Encoded_Bell_State_Mirror_Circuit}. The encoded circuit involves the gates from Ref.~\cite{self2022protecting} with $R_{\rm ZZ}$ and $R_{\rm XX}$ rotation gates having a fixed angle of $+\frac{\pi}{2}$ and $+\frac{3\pi}{2}$, respectively. The total number of gates is $20$ for the version from \cite{self2022protecting} and $180$ using the weakly fault-tolerant Clifford gates presented in our paper. Plots of the probability of an undetectable error and discard rates of the two protocols are shown in Fig.~\ref{fig:Encoded_Bell_State_Mirror_Circuit_Performance_Analysis}. All values were calculated up to second order in the physical error rate $p$ with the methods described in Appendix~\ref{appendix:undetectableErrorCalculation}. As Fig.~\ref{fig:Encoded_Bell_State_Mirror_Circuit_Performance_Analysis} clearly shows, our weakly fault-tolerant gates achieve
better logical error rates at the cost of a larger discard rate for a sufficiently low physical error rate $p$. It is important to note that the above comparison does not consider errors in preparing the initial logical quantum state or measurement and decoding errors. To fully compare the two protocols, simulations (or ideally an experiment) on a trapped ion device should be conducted using the more sophisticated performance tools presented in \cite{self2022protecting}.

We believe our weakly fault-tolerant protocol and the protocol presented by Self, Benedetti and Amaro in \cite{self2022protecting} represent two very promising approaches to early fault-tolerant quantum computation. The main difference between our approaches lies in the trade-off between the probability of an undetectable error and the postselection rate. Both of our papers show that the $[[n,n-2,2]]$ QEDC offers the possibility of incorporating elements of fault tolerance in carrying out quantum computations on current NISQ machines with minimal overhead.

\subsection{Scaling to full fault tolerance}

The motivation underlying this work is straightforward: to improve the performance of near-term quantum computations by adopting some fault-tolerant methods that are within the capabilities of current and near-term quantum processors without other elements whose overhead or complexity is currently too great. These elements are the use of codes, but only high-rate error detection codes; adding ancilla qubits to the encoded gates to catch otherwise undetectable faults, but limiting the number of faults we are guaranteed to find so that we can re-use the ancillas; checking for errors, but only using postselection, not full error correction. For universality we include non-Clifford gates, but do not try to catch analog errors, since that is generally too costly for present machines.

As the capabilities of processors continue to improve, more fault-tolerant methods can be included. One capability that already exists is to do mid-circuit error checks, not just at the end of the computation. As more qubits become available, we can use a larger number of ancillas (rather than re-using them), and go to error-correcting (rather than just detecting) codes. For the present, processing and correcting errors during run-time is still quite difficult, but some correction can be done in classical postprocessing. As error rates come down and conditional operations become faster and more reliable, weakly fault-tolerant non-Clifford gates will become possible, and eventually fully fault-tolerant quantum computation will be achieved.

While these intermediate schemes are not fully fault-tolerant, and therefore not fully scalable, they may bring actual useful quantum computations into reach much sooner than would otherwise be possible. Our approach seems especially promising for quantum simulation \cite{Zhong24}. When combined with error suppression techniques like dynamical decoupling, and other methods like error mitigation, they may allow improved performance without infeasible levels of overhead. In addition, quantum computing architectures such as ion traps and neutral atoms allow for the all-to-all connectivity that we have taken for granted in proving that our gate set is weakly fault-tolerant \cite{schoenberger2024shuttlingscalabletrappedionquantum,reichardt2024logicalcomputationdemonstratedneutral}. As a result, we expect our protocol for universal quantum computation to perform particularly well in implementing near-term quantum computations on these systems.

\section{Conclusions}

This paper has presented weakly fault-tolerant initialization and readout of the QEDC on $n$ qubits, together with a weakly fault-tolerant universal gate set, which yields a complete protocol for quantum computation. Although we cannot correct errors, the $[[n,n-2,2]]$ QEDC and our universal gate set allows us to encode a large number of qubits with a rate approaching 1, and achieve {\it weakly} fault-tolerant quantum computation up to analog errors on our non-Clifford logical $R_Z(\theta)$ gates. The goal of our protocol is to improve performance over circuits with no error detection capability with low additional overhead. This protocol achieves at least an order of magnitude decrease in the probability of an undetectable error in the logical Hadamard and CNOT gates for physical error probabilities lower than $10^{-3}$ for the error model considered. This is primarily beneficial for current and near-term NISQ machines, since calculations are often short, and full quantum error correction is too costly to be beneficial. Our protocol could also be beneficial for quantum computers of the distant future, when gate error rates are so low that weak fault tolerance is sufficient to enable reliable quantum computation.

Although weak fault tolerance allows for universal quantum computation, there is still work that needs to be done to improve it. One of the most important directions for future research is to implement a non-Clifford gate with reduced analog errors that requires little overhead. This would achieve full weak fault tolerance, hopefully without the need for costly magic state distillation. We are sure many other improvements are possible. For example, in this paper we have not studied the trade-offs for mid-circuit stabilizer checks, or other applications of conditional operations. It should be possible to incorporate more elements of true fault tolerance as quantum processors become larger and less noisy, moving to more powerful codes, and approaching scalable quantum computation in the long term while still performing nontrivial smaller computations in the short term.

Our goal for this paper was to present methods for weakly fault-tolerant quantum computation, using a QEDC with a rate approaching 1 as the number of data qubits increases, keeping the overhead low. The protocol we have discussed largely achieves this. Our QEDC and gate set make quantum algorithms that involve a large number of qubits with a short circuit depth more practical on current NISQ machines for error rates that are sufficiently low.

\begin{acknowledgments}
CG and TAB acknowledge many useful conversations with colleagues, especially Rui Chao, Daniel Lidar, Bibek Pokharel, Prithviraj Prabhu, Ben Reichardt, Dawei Zhong and Ken Brown. After this work was essentially complete, but not yet written up, we became aware of the very interesting work in \cite{Yamamoto_2024}, which pursues related ideas and experimentally implements them in an ion trap quantum processor. This work was supported by NSF Grants 1719778, 1911089 and 2316713.
\end{acknowledgments}

\appendix

\section{Details on Circuits For Encoded Clifford Gates in Figures 1--3}
\label{appendix:nonWeaklyFaultTolerantCircuitWalkthrough}

As one can see in the circuit diagrams Figs.~\ref{fig:CNOT}--\ref{fig:Hadamard}, our constructions involve three physical gates. One is the SWAP gate, which is represented as an x on each qubit it operates on with a connecting line drawn between them. The other two are the $R_{\rm ZZ}$ and $R_{\rm XX}$ gate, which are drawn as connected dots with the corresponding label. The placement of the dots indicates which qubits are operated on. The gates in these circuits are the binary symplectic SWAP, $R_{\rm XX}$, and $R_{\rm ZZ}$ gates, which can be found in Eq.~(\ref{eq:binarySymplecticQEDCGates}). The unitary version of the $R_{\rm ZZ}$ and $R_{\rm XX}$ gate can be found in Eqs.~(\ref{ZZ Unitary}) and (\ref{XX Unitary}), respectively. If one is willing to ignore the phase, then it suffices to only use the binary symplectic representation of these two gates to simulate all of the circuits in Figs.~\ref{fig:CNOT}--\ref{fig:Hadamard}. If one instead just replaces the $R_{\rm XX}$ and $R_{\rm ZZ}$ gates in Figs.~\ref{fig:CNOT}--\ref{fig:Hadamard} with their unitary versions, then phase errors may be introduced into the circuit. This is not a major issue, since any error in the phase can be corrected after the circuit through single-qubit operations. For each of these circuits we provide the phase corrections necessary if one implements the circuit with the unitary versions of the $R_{\rm XX}$ and $R_{\rm ZZ}$ gate.

To simulate the action of a circuit on $4$ qubits, we need to expand the SWAP, $R_{\rm ZZ}$, and $R_{\rm XX}$ gate symplectic matrices into $8x8$ matrices. The form of this matrix will depend on which qubits the gate affects. We can then multiply these matrices together using Eq.~(\ref{eq:cliffordSymplecticTransformation}) to find a symplectic matrix that characterizes the entire circuit. To explicitly demonstrate how to fully characterize a quantum circuit in the symplectic formalism, let's work through the circuit shown for the logical Hadamard. Our construction in Fig.~\ref{Hadamardj} operates on $4$ qubits and involves three physical gates. In order, these are $R_{\rm ZZjn}$, $R_{\rm XXj(n-1)}$, and $R_{\rm ZZjn}$. Without loss of generality, we will label these qubits from $1$ to $4$ as in Fig.~\ref{Hadamard1}. As binary matrices operating on $4$ qubits, these gates have the form:
\begin{equation}
\begin{aligned}
\underline{C}_{\rm R_{\rm ZZ14}} & = \begin{pmatrix} 1 & 0 & 0 & 0 & 1 & 0 & 0 & 1 \\ 0 & 1 & 0 & 0 & 0 & 0 & 0 & 0 \\ 0 & 0 & 1 & 0 & 0 & 0 & 0 & 0 \\ 0 & 0 & 0 & 1 & 1 & 0 & 0 & 1 \\ 0 & 0 & 0 & 0 & 1 & 0 & 0 & 0 \\ 0 & 0 & 0 & 0 & 0 & 1 & 0 & 0 \\ 0 & 0 & 0 & 0 & 0 & 0 & 1 & 0 \\ 0 & 0 & 0 & 0 & 0 & 0 & 0 & 1 \end{pmatrix} , \\ \rule{0pt}{15ex} 
\underline{C}_{\rm R_{\rm XX13}} & = \begin{pmatrix} 1 & 0 & 0 & 0 & 0 & 0 & 0 & 0 \\ 0 & 1 & 0 & 0 & 0 & 0 & 0 & 0 \\ 0 & 0 & 1 & 0 & 0 & 0 & 0 & 0 \\ 0 & 0 & 0 & 1 & 0 & 0 & 0 & 0 \\ 1 & 0 & 1 & 0 & 1 & 0 & 0 & 0 \\ 0 & 0 & 0 & 0 & 0 & 1 & 0 & 0 \\ 1 & 0 & 1 & 0 & 0 & 0 & 1 & 0 \\ 0 & 0 & 0 & 0 & 0 & 0 & 0 & 1 \end{pmatrix} .
\end{aligned}
\end{equation}


To find the symplectic matrix that characterizes the entire circuit, we simply apply Eq.~(\ref{eq:cliffordSymplecticTransformation}) two times, with the first $R_{\rm ZZ}$ gate acting as our initial matrix of generators, which yields:
\begin{equation}
\underline{C}_{\rm Hadamard_{1}} = \begin{pmatrix} 0 & 0 & 1 & 0 & 1 & 0 & 0 & 1 \\ 0 & 1 & 0 & 0 & 0 & 0 & 0 & 0 \\ 0 & 0 & 1 & 0 & 0 & 0 & 0 & 0 \\ 1 & 0 & 1 & 1 & 1 & 0 & 0 & 1 \\ 1 & 0 & 1 & 0 & 0 & 0 & 0 & 1 \\ 0 & 0 & 0 & 0 & 0 & 1 & 0 & 0 \\ 1 & 0 & 1 & 0 & 1 & 0 & 1 & 1 \\ 0 & 0 & 0 & 0 & 0 & 0 & 0 & 1 \end{pmatrix}
\end{equation}


To confirm that our logical gate is in fact a Hadamard, we can track how the logical operators and stabilizer generators of the $[[n,n-2,2]]$ code are transformed by the circuit. Explicitly, the logical operators transform in the following manner:
\begin{eqnarray}
{\bar{X}_1} &=& {XIXI} \rightarrow {YIXZ} \rightarrow {ZIIZ} \rightarrow {ZIIZ} = {\bar{Z}_1} \nonumber \\
{\bar{Z}_1} &=& {ZIIZ} \rightarrow {ZIIZ} \rightarrow {YIXZ} \rightarrow {XIXI} = {\bar{X}_1} \nonumber\\
{\bar{X}_2} &=& {IXXI} \rightarrow {IXXI} \rightarrow {IXXI} \rightarrow {IXXI} = {\bar{X}_2} \nonumber\\
{\bar{Z}_2} &=& {IZIZ} \rightarrow {IZIZ} \rightarrow {IZIZ} \rightarrow {IZIZ} = {\bar{Z}_2}
\label{eq:logicalOperatorsEvolution}
\end{eqnarray}

We also need to check that the circuit does not change the stabilizer generators of our code. As one can quickly show, our circuit transforms the stabilizer generators as follows:
\begin{eqnarray}
{XXXX} &\rightarrow& {XXXX} \rightarrow {XXXX} \rightarrow {XXXX} \nonumber\\
{ZZZZ} &\rightarrow& {ZZZZ} \rightarrow {ZZZZ} \rightarrow {ZZZZ}
\label{eq:stabilizerGeneratorsEvolution} 
\end{eqnarray}

We have confirmed that our circuit implements a logical Hadamard on the $j$th qubit, which we have labeled qubit $1$. We now need to check how it affects the phase of our logical operators and stabilizer generators. One way to do this is to determine how the $R_{\rm XX}$ and $R_{\rm ZZ}$ gates affect the phase of an operator. We will make use of the unitary representations of these gates. From Eqs.~(\ref{ZZ Unitary}) and (\ref{XX Unitary}), one can immediately prove through unitary evolution that the $R_{\rm ZZ}$ and $R_{\rm XX}$ gates implement the following phase transformations:
\begin{eqnarray}
{+XI} &\rightarrow& {+YZ}  \nonumber\\
{+ZI} &\rightarrow& {+YX}  \nonumber\\
{+YX} &\rightarrow& {-ZI}  \nonumber\\
{+YZ} &\rightarrow& {-XI}
\label{eq:phaseTransformations}
\end{eqnarray}
Using these results, we can track the phase of the logical operators and stabilizer generators to show that the logical Hadamard presented in Fig.~\ref{Hadamardj} does in fact produce a $-1$ phase error:
\begin{eqnarray}
\begin{array}{rccl}
{+\bar{X}_1} &=& {+XIXI} \rightarrow {+YIXZ} \rightarrow {-ZIIZ} & \rightarrow {-ZIIZ} \\
& & &= {-\bar{Z}_1} \nonumber \\
{+\bar{Z}_1} &=& {+ZIIZ} \rightarrow {+ZIIZ} \rightarrow {+YIXZ} & \rightarrow {-XIXI} \\
& & &= {-\bar{X}_1}
\end{array}
\end{eqnarray}

We can correct this phase error with the Pauli operators $Z_j Z_n$ and $X_j X_{n-1}$. Our final step is to see if our construction for the logical Hadamard is weakly fault-tolerant. To analyze our depolarizing error model, it is sufficient to simply insert Pauli errors after each gate and then check how they propagate through the circuit. Although in practice our entire quantum circuit may be composed of more elements, it is sufficient to check the form of errors at the end of our specific circuit construction. Any error that commutes with the stabilizer generators of the code is an undetectable error. As an explicit example, lets see how a ZX error after the first $R_{\rm ZZ}$ gate propagates:
\begin{equation}
{ZIIX} \rightarrow {YIXX} \rightarrow {YIXX} .
\end{equation}
This error is detectable since it anticommutes with the all-$X$ stabilizer. However, if we had a $R_{\rm ZZ}$ error after the first gate instead,
\begin{equation}
{ZIIZ} \rightarrow {YIXZ} \rightarrow {XIXI} .
\end{equation}
An $XIXI$ error commutes with all of the stabilizers of the QEDC and is thus undetectable. A single faulty gate in this circuit can produce a Pauli error that is undetectable, so we can conclude that the construction in Fig.~\ref{Hadamardj} is not weakly fault-tolerant! Through the exact same analysis, one can work through all of the circuits present in Figs.~\ref{fig:CNOT}--\ref{fig:Hadamard}.

\section{Details of the Weakly Fault-Tolerant Circuits in Figures 4--5}
\label{appendix:WeaklyFaultTolerantCircuitWalkthrough}

In a manner similar to Appendix~\ref{appendix:nonWeaklyFaultTolerantCircuitWalkthrough}, one can fully characterize the circuit and check if it is weakly fault-tolerant. We will partially work through one of the circuits, although there are too many errors to keep track of by hand. To prove that these circuits are weakly fault-tolerant, we used a Mathematica script to generate errors and check that a single faulty gate cannot generate a Pauli error that becomes undetectable at the end of the circuit. The general outline of how to do this is as follows. For each gate in the circuit, we associate a set of Pauli errors that can then be inserted directly after it. For single-qubit gates, this set comprises the Pauli errors $X$, $Y$, and $Z$. To extend this to two-qubit gates, simply take all possible combinations of these errors (and $I$) over two qubits. This generates a set of $15$ nontrivial Pauli errors. To check for weak fault tolerance in a circuit, it is sufficient to exhaustively check every possible Pauli error a single faulty gate can produce, evolve the error to the end of the circuit, and then check if it anticommutes with one of the stabilizer generators of the QEDC or the ancillas. If all of these errors that occur with a probability of $O(p)$ are detectable, then we say that the circuit is weakly fault-tolerant.

To make the analysis more concrete, let's look at Fig.~\ref{ZZ,Phi+}. To show that this circuit in fact implements a $R_{\rm ZZ}$ gate, it is sufficient to carry out matrix multiplication of all the gates, as discussed in Appendix~\ref{appendix:nonWeaklyFaultTolerantCircuitWalkthrough}, and show that its matrix representation is equivalent to the $R_{\rm ZZ}$ gate up to a stabilizer generator on the ancillas. By ``up to a stabilizer generator'', we mean that any deviation between the weakly fault-tolerant $R_{\rm ZZ}$ gate and the original $R_{\rm ZZ}$ gate must be equivalent to multiplying by a stabilizer generator on the ancilla qubits. So, for example, our weakly fault-tolerant circuit in Fig.~\ref{fig:FTZZ}, which we will denote as $\underline{C}_{\rm R_{\rm ZZ},\ket{\Phi+}}$, implements the following transformation:
\begin{equation}
\underline{C}_{\rm R_{\rm ZZ}, \ket{\Phi+}} = \begin{pmatrix} 0 & 0 & 0 & 0 & 1 & 1 & 1 & 1 \\ 1 & 1 & 0 & 0 & 1 & 1 & 1 & 1 \\ 1 & 0 & 1 & 0 & 0 & 0 & 1 & 1 \\ 1 & 0 & 0 & 1 & 0 & 0 & 1 & 1 \\ 1 & 0 & 0 & 0 & 1 & 0 & 1 & 1 \\ 1 & 1 & 0 & 0 & 1 & 0 & 1 & 1 \\ 0 & 0 & 0 & 0 & 0 & 0 & 1 & 0 \\ 0 & 0 & 0 & 0 & 0 & 0 & 0 & 1 \end{pmatrix} .
\end{equation}
Although we could keep track of the stabilizer generators and data qubit operators as in Eqs.~(\ref{eq:logicalOperatorsEvolution}--\ref{eq:stabilizerGeneratorsEvolution}), one can instead gather these generators into a matrix and track their evolution through Eq.~(\ref{eq:cliffordSymplecticTransformation}). Without loss of generality, we will label our ancillas as qubits $1$ and $2$ and our data qubits as $3$ and $4$. For Fig.~\ref{ZZ,Phi+}, our starting matrix of generators is thus
\begin{equation}
\underline{M} = \begin{pmatrix} 1 & 1 & 0 & 0 & 0 & 0 & 0 & 0 \\ 0 & 0 & 0 & 0 & 1 & 1 & 0 & 0 \\ 0 & 0 & 1 & 0 & 0 & 0 & 0 & 0 \\ 0 & 0 & 0 & 1 & 0 & 0 & 0 & 0 \\ 0 & 0 & 0 & 0 & 0 & 0 & 1 & 0 \\ 0 & 0 & 0 & 0 & 0 & 0 & 0 & 1 \end{pmatrix} .
\end{equation}
The first two rows of $\underline{M}$ are the stabilizer generators of the $\ket{\Phi+}$ state. The other $4$ rows are the single-qubit $X$ and $Z$ operators on our data qubits. Applying Eq.~(\ref{eq:cliffordSymplecticTransformation}) yields
\begin{equation}
\begin{aligned}
\underline{M}\underline{C}_{\rm R_{\rm ZZ}, \ket{\Phi+}} = & \begin{pmatrix} 1 & 1 & 0 & 0 & 0 & 0 & 0 & 0 \\ 0 & 1 & 0 & 0 & 0 & 0 & 0 & 0 \\ 1 & 0 & 1 & 0 & 0 & 0 & 1 & 1 \\ 1 & 0 & 0 & 1 & 0 & 0 & 1 & 1 \\ 0 & 0 & 0 & 0 & 0 & 0 & 1 & 0 \\ 0 & 0 & 0 & 0 & 0 & 0 & 0 & 1 \end{pmatrix} \\ \rule{0pt}{11.5ex} 
\rightarrow & \begin{pmatrix} 0 & 0 & 0 & 0 & 0 & 0 & 0 & 0 \\ 0 & 0 & 0 & 0 & 0 & 0 & 0 & 0 \\ 0 & 0 & 1 & 0 & 0 & 0 & 1 & 1 \\ 0 & 0 & 0 & 1 & 0 & 0 & 1 & 1 \\ 0 & 0 & 0 & 0 & 0 & 0 & 1 & 0 \\ 0 & 0 & 0 & 0 & 0 & 0 & 0 & 1 \end{pmatrix} ,
\end{aligned}
\end{equation}
where we have projected out the stabilizer generators of the $\ket{++}$ state. To show that the circuit in Fig.~\ref{ZZ,Phi+} is weakly fault-tolerant, we can exhaustively generate every possible single-gate Pauli error and evolve each of them to the end of the circuit. We then just need to check that all of these errors anticommute with either the stabilizer generators of the $\ket{++}$ state (the ancilla) or one of the $XX$ and $ZZ$ operators on qubits $3$ and $4$. These last two operators represent the portion of the stabilizer generators of the $[[n,n-2,2]]$ QEDC present on the data qubits. Explicitly, we have $15$ possible Pauli errors for each two-qubit gate and $3$ for our lone single-qubit gate. One then shows that this construction is weakly fault-tolerant by confirming that all $123$ possible single-faulty-gate errors are detectable at the end of the circuit. As we show in Sec.~\ref{sec:weaklyFaultTolerant}, these errors will remain detectable throughout a much larger encoded circuit where for the purposes of showing weak fault tolerance we can consider all future gate operations error-free. Carrying out this same procedure for the other $3$ weakly fault-tolerant circuits will show that they all implement the expected $R_{\rm ZZ}$ or $R_{\rm XX}$ gate on the data qubits in a weakly fault-tolerant manner. Finally, we need to consider how these weakly fault-tolerant gates affect the phase. Through the same procedure outlined in Appendix~\ref{appendix:nonWeaklyFaultTolerantCircuitWalkthrough}, one can keep track of the phase and determine which generators have their phases incorrectly flipped. Applying our recovery operations in Table~\ref{tab:phaseCorrection} will fix each of these errors. This completes the analysis of our weakly fault-tolerant constructions for the $R_{\rm ZZ}$ and $R_{\rm XX}$ gates.

\section{Details on Calculating The Probability of an Undetectable Error}
\label{appendix:undetectableErrorCalculation}

As mentioned earlier, let us denote the probability of a single gate error as $p$ (the physical error rate of our gates). This means that the probably of no error is $1-p$ and the probability of any single possible Pauli error on a single two-qubit gate is ${p}/{15}$. Our program tells us how many undetectable errors there are at a given order of error. For the non-weakly fault-tolerant circuits, the total possible number of errors at this order is a simple binomial distribution $\binom{n}{k}$ multiplied by the total number of Pauli errors $15^k$ where $n$ represents the total number of two-qubit gates in our circuit and $k$ is the error order we are considering. From the number of undetectable errors we can get the proportion of errors that are detectable, which is a number between $0$ and $1$. Simply multiplying this fraction by $\binom{n}{k}(1-p)^{n-k}(p)^{k}$ gives the probability of a detectable error at this order. Specifically, we can rewrite the probability that $k$ gates fail and cause a detectable error using conditional probability. $\binom{n}{k}(1-p)^{n-k}(p)^{k}$ is equivalent to the probability that $k$ gates fail. The proportion of $k$ errors that are detectable gives the conditional probability that an error is detectable given $k$ gate failures occurred, since we assume that the probability of a gate failure is independent and identically distributed (i.i.d.) for all two-qubit gates. For weakly fault-tolerant circuits, the total number of errors at any order is slightly more difficult to calculate due to the presence of single-qubit gates, which only have $3$ possible Pauli errors. In order to not have to keep track of which gates caused the undetectable error, we only consider the effect of these single-qubit gates when calculating what proportion of errors are detectable. We then just multiply this by $\binom{n}{k}(1-p)^{n-k}(p)^{k}$ to get the probability of a detectable error at order $k$. The error due to this approximation depends on the distribution of undetectable errors. The probability of any single undetectable error of $O(p^2)$ and higher is already extremely small for the physical error probabilities we are considering and only ${1}/{9}$ of our physical gates are single-qubit gates. Moreover, single-qubit gates have much lower error rates when compared to two-qubit gates. Due to this, any approximation error will have a negligible impact on the graphs presented in the error analysis section.

As a concrete example, let's consider the analysis for errors of second order for the weakly fault-tolerant encoded Hadamard gate. In this case, there are $3,108$ undetectable errors as given by our program and $62,259$ detectable errors. The probability of a detectable error at this order is thus
\begin{equation}
\binom{27}{2}*\frac{62259}{65367}*(1-p)^{25}(p)^{2},
\end{equation}
where $65,367$ is the exact number of possible Pauli errors of $O(p^2)$. Carrying out a similar calculation for errors of order zero (no error), first order (all errors at this level will be detectable), and third order, summing them together, and then subtracting them from unity will give the curve for the probability of an undetectable error for the weakly fault-tolerant Hadamard gate.

\bibliography{References.bib}

\end{document}